\documentclass[twoside,twocolumn,english,aps,amsmath,amssymb,reprint,pra,superscriptaddress]{revtex4-1}
\usepackage[T1]{fontenc}
\usepackage[latin9]{inputenc}
\usepackage{color}
\usepackage{amsmath}
\usepackage{amssymb}
\usepackage{graphicx}

\makeatletter

\providecommand{\tabularnewline}{\\}

\usepackage{babel}
\usepackage{babel}

\makeatother

\usepackage{babel}
\begin{document}

\title{Aggregating Quantum Networks}

\author{Nicolò Lo Piparo}
\email{nicopale@gmail.com}

\affiliation{National Institute of Informatics, 2-1-2 Hitotsubashi, Chiyoda-ku,
Tokyo 101-8430, Japan.}

\author{Michael Hanks}

\affiliation{National Institute of Informatics, 2-1-2 Hitotsubashi, Chiyoda-ku,
Tokyo 101-8430, Japan.}

\author{Kae Nemoto}

\affiliation{National Institute of Informatics, 2-1-2 Hitotsubashi, Chiyoda-ku,
Tokyo 101-8430, Japan.}

\date{\today}

\author{William J. Munro}

\affiliation{NTT Basic Research Laboratories \& NTT Research Center for Theoretical
Quantum Physics, NTT Corporation, 3-1 Morinosato-Wakamiya, Atsugi,
Kanagawa, 243-0198, Japan.}

\affiliation{National Institute of Informatics, 2-1-2 Hitotsubashi, Chiyoda-ku,
Tokyo 101-8430, Japan.}

\date{\today}
\begin{abstract}
Quantum networking allows the transmission of information in ways
unavailable in the classical world. Single packets of information
can now be split and transmitted in a coherent way over different
routes. This aggregation allows information to be transmitted in a
fault tolerant way between different parts of the quantum network
(or the future internet) \textendash{} even when that is not achievable
with a single path approach. It is a quantum phenomenon not available
in conventional telecommunication networks either. We show how the
multiplexing of independent quantum channels allows a distributed
form of quantum error correction to protect \textcolor{black}{the
transmission of} quantum information between nodes or users of a quantum
network. Combined with spatial-temporal single photon multiplexing
we observe a significant drop in network resources required to transmit
that quantum signal \textendash{} even when only two channels are
involved. This work goes far beyond the concepts of channel capacities
and shows how quantum networking may operate in the future. Further
it shows that quantum networks are likely to operate differently from
their classical counterparts which is an important distinction as
we design larger scale ones.
\end{abstract}
\maketitle

\section{Introduction}

Our recent advances have brought quantum technologies from an abstract
thought experiment to reality with many pivotal demonstrations being
achieved and even devices available for commercial use \cite{Qtech,Qtech2,QDevices1,QDevices2,QDevices3,QDevices4,QDevices5}.
Such technologies can be characterized into a number of broad areas
including quantum computation \cite{Qcomp1,Qcomp2,Quantum_comp2,Quantum_comp3,Quantum_comp4},
communication \cite{QComm,QC_Bill,Quantumcomm,DLCZ_p,QKD02,QKD03,QKD1,QKD2000,QKD_transmission},
metrology \cite{Qmetrology1,Qmetrology2}, sensing and imaging \cite{QSensing,QImaging3,QImaging2}.
We have achieved remarkable control over these systems and recently
a ``quantum advantage'' has been achieved in the quantum computational
regime on a monolithic chip \cite{QC_google}. Quantum communication
also holds significant promise \textendash{} but clearly is not as
advanced yet. We have however seen quantum key distribution networks
operating on a continental scale \cite{QKD2000,satelliteQKD} but
this is far from a future multi-purpose global quantum internet \cite{Q_internet}.

In any future quantum internet, we are going to need to be able to
send (or transmit) quantum information over large distances \textendash{}
potentially through many intermediate routing nodes. Whether this
occurs by the direct transmission of such information \cite{QKD_transmission,Bill_nature_phot}
or by quantum teleportation \cite{tel_1,tel_2,Qtel2,Qtel3} after
an entangled resource has been established \cite{tel_network} we
already know that some mechanism to handle both loss and local gate
errors will be necessary \cite{Qcomp1}. Approaches based on quantum
error detection codes \cite{QP2Deutsch,Ent_pur}, while attractive
in the short term, are\textcolor{black}{{} performance }limited due
to their reliance on probabilistic operations \cite{QP2Deutsch,Bill_nature_phot,Bill_2}.
Scheme using quantum error correction for both loss and gate errors
overcome such issues but are technically much more challenging \cite{redundancy_code,Reed_Sal,surfacecode,GKP1,Bill_nature_phot,Bill_3,cat_states,binomial_code}.
Still for large scale quantum networks, they are likely to be the
only viable approach.

We can picture a general quantum network as a complex network involving
certain nodes connected to each other by quantum links. Within these
nodes we have a certain number of quantum bits associated with those
links to the adjacent node. In any multiuser scenario we are likely
to be in a constrained resource situation and there may not be enough
resources associated with one path to that adjacent nodes at that
time to allow the reliable transmission of one\textquoteright s error
correction encoded signal. However, in these complex networks there
are likely to be multiple paths between nodes (some potentially with
immediate nodes in between). The natural question that arises is:
\textit{if }\textit{\textcolor{black}{no}}\textit{ path has sufficient
resources by itself (either the number of qubits with the node or
the capacity of the channel itself), can we combine (\textquoteleft aggregate\textquoteright )
them together to achieve it?} This will be the focus of our paper. 

The concept of quantum network aggregation is of course not new having
been considered by a number of groups in the context of establishing
bounds of the quantum and private capacities \cite{quantum_capt,q_capacity3,q_capacity2,stefano1,stefano2,q_capacity1}.
Unfortunately, those capacities tell us limited information about
what would have in a dynamical network where resources are being continuously
utilized and potentially consumed, nor do they tell us how such aggregation
can take place. In our work here we consider the simple situation
of two resource constrained parties (Alice and Bob) who have two independent
channels between themselves with different transmission properties.
We show that single quantum states (packets) can be transmitted simultaneously
over both channels in a coherent fashion \textendash{} enabling its
transmission which would not have been possible by either individual
channel along. Further this adds a new capability to networking not
present in the telecommunications world.
\begin{figure*}
\begin{centering}
\includegraphics[scale=0.25]{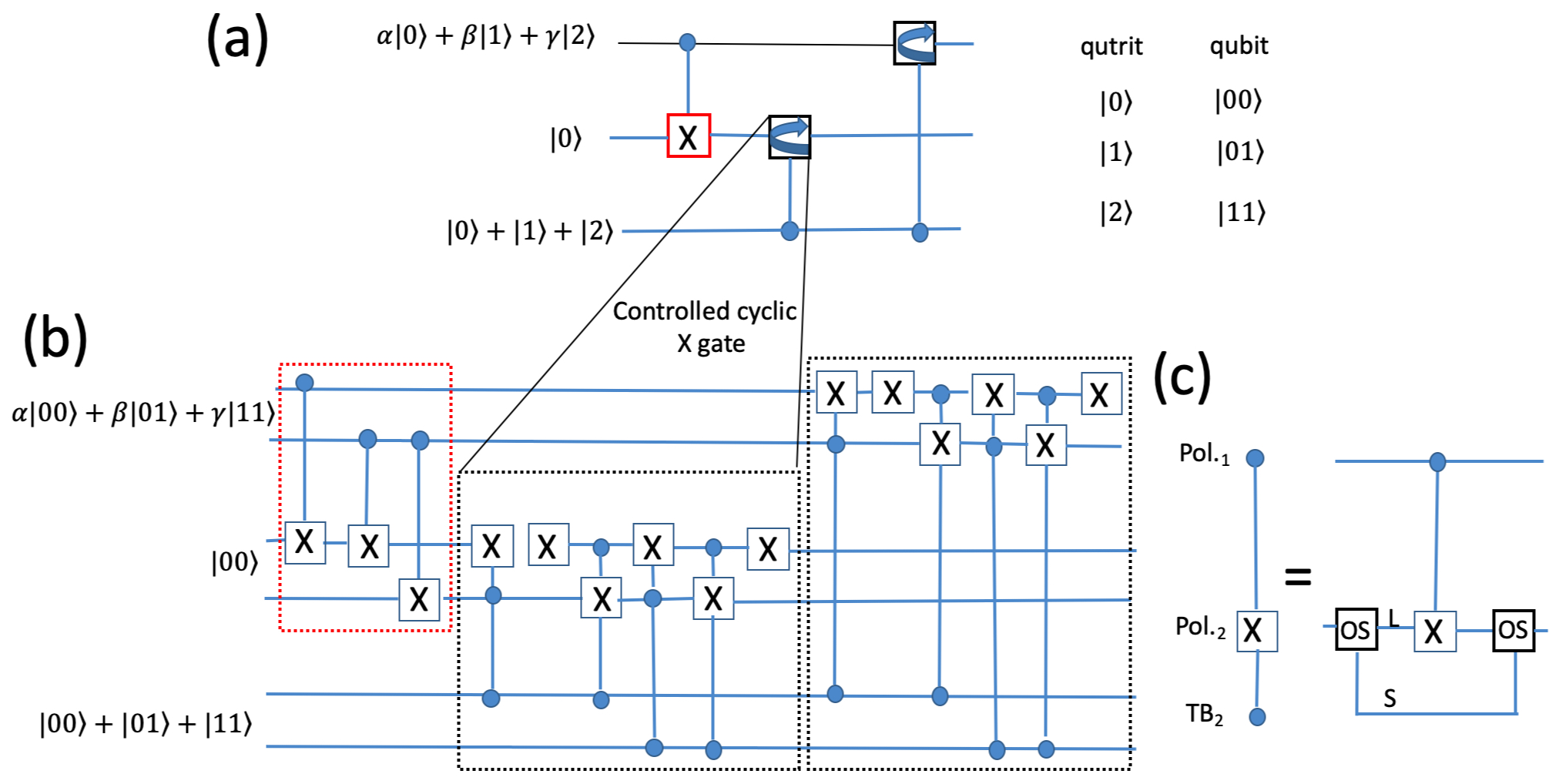}
\par\end{centering}
\caption{\label{fig:logic_qutrit_qubit}(a) Logic circuit used for the encoding
of the (3,1,2) QRS code with qutrits, (b) gives the corresponding
circuit using qubits while (c) illustrates how the Toffoli gate can
be implemented for multiplexed photons using only a single photon-photon
CNOT gate. This is potentially an important resource saving \cite{toffoli1,toffoli2}.
Here Pol.$_{1}$, Pol.$_{2}$ and TB$_{2}$ stand for the polarization
DOFs of photon 1, photon 2 and time-bin DOF of photon 2 respectively.
OS is an optical switch that splits into two other modes (Long and
Short) photon 2. A CNOT between photon 1 and the Long component of
photon 2 is performed followed by another OS that recombines the modes.}
\end{figure*}

Recently, it been found that quantum multiplexing \cite{QuM} \textendash{}
the process whereby information is encoded into different degrees
of freedom of a photon \cite{hyperentanglement1,hyper2,hyper3} \textendash{}
can dramatically decrease both the number of qubits required within
the node and also the number of photons being transmitted through
the channel (still this may not be enough). However, in conjunction
with aggregation those advantages may be enhanced further.

The paper is divided as following: In Section II we briefly introduce
communication-based quantum error correction in terms of the quantum
Reed-Solomon (QRS) code and show how quantum multiplexing can be applied
to it. Then in Section III we illustrate our aggregation scheme between
two nodes with two different channels connecting them. This is followed
up in Section IV with the extension to spatial-temporal single photon
multiplexing. We conclude in Section V. 

\section{The quantum Reed-Solomon code}

It is well known that advanced quantum repeaters will be based on
quantum error correction techniques \cite{Bill_nature_phot}. There
are a huge variety of codes available for use in these circumstances
\cite{redundancy_code,Reed_Sal,surfacecode,GKP1} but here our primary
focus will be on the Quantum Reed Solomon (QRS) code \cite{Reed_Sal}.
Such a code has excellent properties to handle channel loss events
and has also recently been shown that the number of physical resources
required to implement it can be dramatically reduced by using a quantum
multiplexing approach \cite{QuM}. The QRS code has been used in several
applications in the last few years \cite{Reed_Sal}. In this Section
we will review the QRS code and how it can be used in the quantum
multiplexing regime. 

\noindent \textcolor{black}{The QRS code is typically written in the
form $[[d,2k-d,d-k+1]]_{d}$, where $2k-d$ logic qudits of dimension
$d$ (prime number) are encoded into $d$ physical qudits, in such
a way that when $d-k$ or less qudits are lost the encoded qudit can
be retrieved. A simple example of such a code is the $[[3,1,2]]_{3}$
code. Here a logic qutrit $\left.|D\right\rangle $ is encoded using
three physical qutrits as \cite{PS_QRS}:}

\begin{equation}
\left.|D\right\rangle =\left.\alpha|0\right\rangle _{L}+\left.\beta|1\right\rangle _{L}+\left.\gamma|2\right\rangle _{L},\label{eq:logic_qutrit}
\end{equation}
where (omitting the nor\textcolor{black}{malization constants for
simplicity) $\left.|0\right\rangle _{L}=\left.|000\right\rangle +\left.|111\right\rangle +\left.|222\right\rangle ,$
$\left.|1\right\rangle _{L}=\left.|012\right\rangle +\left.|120\right\rangle +\left.|201\right\rangle $
and $\left.|2\right\rangle _{L}=\left.|021\right\rangle +\left.|102\right\rangle +\left.|210\right\rangle $
(see Fig. \ref{fig:logic_qutrit_qubit}(a) and Appendix A for further
details). This code has used physical qutrits in its encoding, however
it may be more convenient to use physical qubits due to their more
common nature. As such we can use two qubits to encode one physical
qutrit meaning six qubits are needed for the logical state to encode
the $[[3,1,2]]_{3}$ code (see Fig. \ref{fig:logic_qutrit_qubit}(b)
and Appendix B). One can extend this encoding procedure to the multiplexed
case \cite{QuM} in which each photon is carrying two qubits of information.
This approach reduces number of physical resources in an error correction
scheme \cite{QuM} (the usual Toffoli gate reduces to a simple CNOT
gate as shown in Fig. \ref{fig:logic_qutrit_qubit}(c)). F}urther
details on how the logic qutrit $\left.|D\right\rangle $ can be created
with the quantum multiplexing approach is shown in Appendix C. 

In using these error correction techniques it is important to consider
our figures of merit for how we assess the performance of those approaches.
The main figure of merit we are interested in our analysis is the
probability of successfully transmitting \textcolor{black}{$d$ qudits
over a lossy channel, $P_{S}(1\mathrm{ch)}$. For that generic $[[d,2k-d,d-k+1]]_{d}$
QRS code, in which photons are carrying $q$ qubits of information,
any qudit depends on the successful transmission of $\lceil\log_{2}(d)/q\rceil$
photons \cite{Reed_Sal}. Therefore $P_{S}(1\mathrm{ch)}$ is given
by \cite{PS_QRS}:}

\noindent 
\begin{equation}
P_{S}(1\mathrm{ch)}=\sum_{j=0}^{d-k}{d \choose {j}}P_{t_{1}}^{(d-j)}(1-P_{t_{1}})^{j},\label{eq:succ_prob_single_ch}
\end{equation}
where $P_{t_{1}}=p_{t_{1}}^{\lceil\frac{\log_{2}(d)}{q}\rceil}$ is
the transmission probability of a qudit whose photons are traveling
in the channel with transmission probability $p_{t_{1}.}$\textcolor{black}{{}
This approach assumes that all photons are transmitted over identical
channels following the same route and that the number of qudits is
a prime number. However, one can also think of encoding the initial
state in a prime number $d$ of qudits but sending only $d-l$ qudits,
providing that the transmission channel has higher transmission probability.
For a certain $\bar{l,}$ $d-\bar{l}$ will be a prime number and
the resulting $d-\bar{l}$ QRS code in which $d$ qudits are initially
encoded but $d-\bar{l}$ only have been sent, requires, obviously,
a channel with a slightly higher transmission probability compared
to the case in which $d-\bar{l}$ qudits have been encoded and sent
(the usual QRS code). Therefore, this approach does not show any advantages
in terms of reduction of physical resources as well as of allowing
a worse channel capacity. However, for such a generic $d-l$ QRS code
one may think of using the $l$ encoded qudits, which have been not
sent, for other purposes. In the next Section we investigate the case
in which these $l$ encoded qudits are also sent to Bob in a quantum
channel having a different transmission probability than the one in
which $d-l$ are traveling.}

\section{Simple quantum network aggregation - I }

\begin{figure}
\begin{centering}
\includegraphics[scale=0.44]{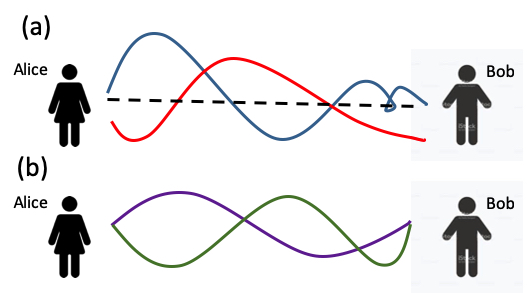}
\par\end{centering}
\caption{\label{fig:Example of aggregate network}In (a), three quantum channels
connecting two parties (Alice and Bob) with different number of qudits
and transmission probabilities required to reach the threshold success
probability $\overline{P}_{S}$ for the 7-qudits QRS code. In (b),
the same threshold probability can be reached with the aggregating
network approach in which Alice and Bob are now connected by two channels
having less transmission probability than the blue and red channel
or less qudits per channel than the blue and black channel. }
\end{figure}

In this Section we will describe the general application of our aggregate
network approach to the QRS code and illustrate a possible advantage
that arises. Let us assume that two parties (Alice and Bob) want to
exchange several quantum bits of information by encoding them into
a logic state using the QRS code. There are several ways they can
achieve this. Figure \ref{fig:Example of aggregate network}(a) shows
Alice and Bob connected by a number of channels, each of them having
different capacities for carrying a specific number of qudits and
transmission probabilities. It is important at this stage to establish
a success probability threshold $\overline{P}_{S}$ which will be
used to quantify the quality of transmitted quantum information. This
we set at $\overline{P}_{S}=0.995$ a value normally associated with
fault-tolerant quantum computation \cite{ftolerant_threshold}.

More specifically consider that Alice has available to her, 3 qudits
that may travel over the blue channel with transmission probability
$P_{t_{blue}}=0.98,$ 5 qudits that may travel over the red channel
with $P_{t_{red}}=0.972,$ and 7 qudits for the black channel with
$P_{t_{black}}=0.96.$ Each of these channels with their complement
of qudits is just sufficient to reach the target threshold pro\textcolor{black}{bability
(we use Eq. \ref{eq:succ_prob_single_ch}).} However, this is the
situation in which she uses each channel independently of each other.
She could however combine these channels together using 5 qudits,
in which 2 of them are in the blue channel, 2 in the red channel and
one qudit in the black channel. This combination meets the threshold
but uses less qudits per channel than the no aggregation technique. 

The above example indicates that aggregating channels of different
quality together can allow one to use a mixture of resources and optimize
their usage. For instance, using 2 blue channel qudits and 3 qudits
in a channel with lower transmission probability allows one to reach
the threshold success probability without consuming all the resources
in one channel. This is particularly important in a multiuser scenario.
Our considerations using this aggregation strategy also allows us
to potentially decrease the transmission probabilities on each channel.
We can combine two channels (see Fig. \ref{fig:Example of aggregate network}(b))
where 3 qudits (4 qudits) are transmitted through lossy channels with
probability $P_{t_{purple}}=0.97(P_{t_{green}}=0.95)$ and still reach
the threshold success probability. This is interesting because it
gives us another tool we can use in our network aggregation. 
\begin{figure*}
\begin{centering}
\includegraphics[width=1\textwidth]{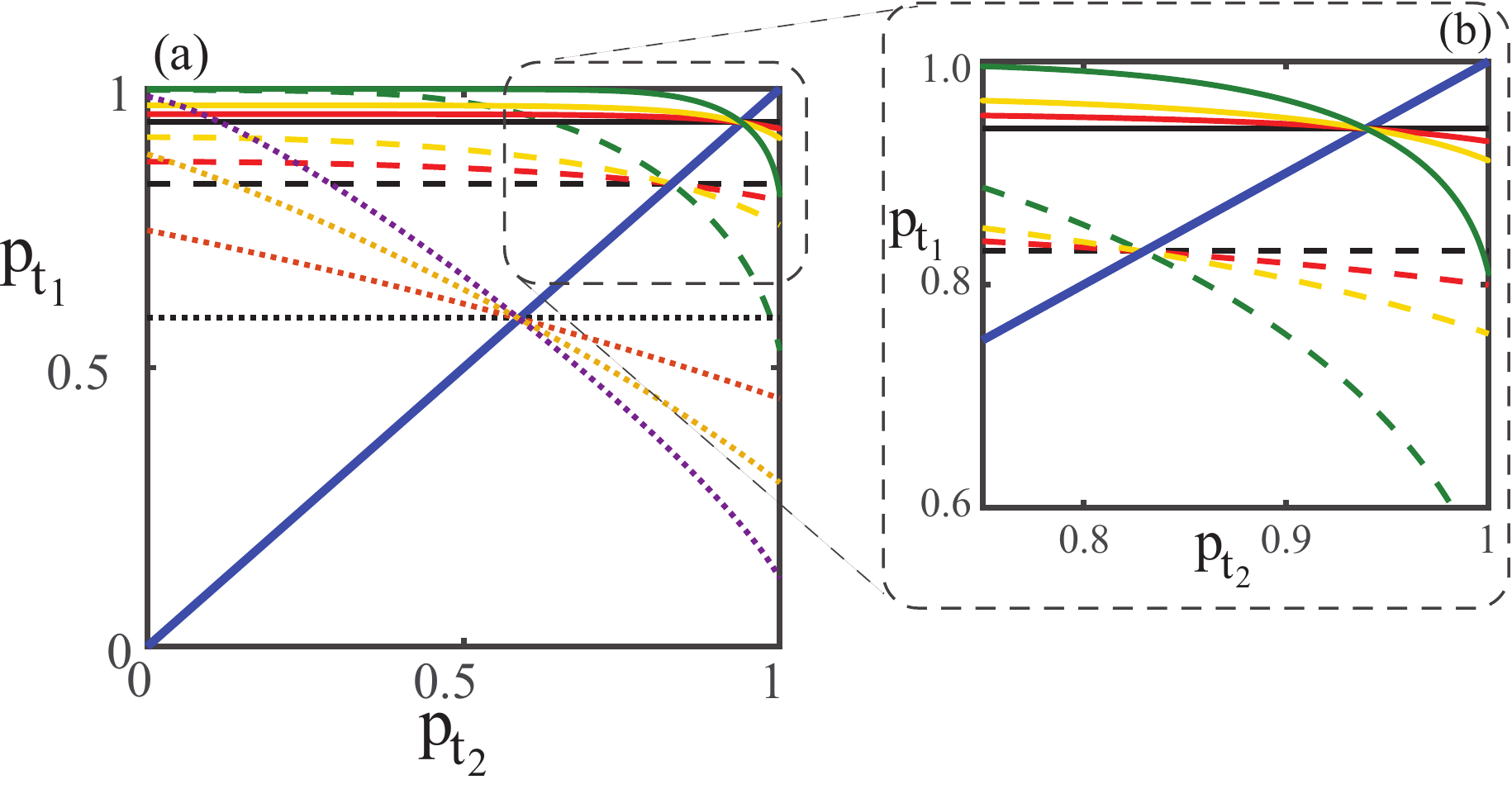}
\par\end{centering}
\caption{\label{fig:first_level_graph}(a) Photon transmission probabilities
required to reach $\overline{P}_{S}$ for different QRS codes. The
dotted curves correspond to the quantum multiplexing 211-qudit QRS
case, in which each photon is carrying 8 qubits, where all qudits
are traveling in the same channel (black), 50 (orange), 80 (yellow)
and 100 (purple) qudits are traveling in the channel having transmission
probability $p_{t_{2}}.$ The solid (dashed) curves refer to the no
multiplexing (multiplexing) case for the 43 qudits QRS code. The black
curves are the cases when all qudits are traveling in the same channel.
The red, yellow and green curves are, respectively, the cases when
5, 10 and 20 qudits are traveling in the channel with transmission
probability $p_{t_{2}}.$ In (b), a zoom of (a) that highlights the
advantage of using the aggregation network approach for a range of
values of the transmission probabilities of the 43-qudit QRS code.}
\end{figure*}

A natural question that arises now is what the overall success probability
when multiple channels are used in an aggregative fashion. It is illustrative
here to explore the 2 channels example. Here we will assume that all
photons are encoding a specific qudit and propagate in one of these
2 channels. In this case, the overall success probability is

\begin{equation}
\begin{array}{c}
P_{S}\left(2\mathrm{ch}\right)=\sum_{i=0}^{(d-1)/2}\sum_{j=0}^{i}{n \choose {i-j}}P_{t_{2}}^{n-(i-j)}(1-P_{t_{2}})^{i-j}\\
\times{d-n \choose {j}}P_{t_{1}}^{d-n-j}(1-P_{t_{1}})^{j},
\end{array}\label{eq:succ_prob_double_ch}
\end{equation}
where $P_{t_{2}}=p_{t_{2}}^{\lceil\frac{\log_{2}(d)}{q}\rceil}$ is
the transmission probability of all qudits encoded with photons traveling
in the lossy channel with transmission probability $p_{t_{2}}.$ This
is valid for both the multiplexed and unmultiplexed cases.

The example we have explored so far uses few qudits and requires rather
high transmission probabilities. To move to more realistic situations
in which the transmission probabilities can be much lower we need
to increase the size the error correction code. The 43 qudit QRS error
correcting code is a nice compromise here as it allows us to explore
situations in which, by increasing the quality of one channel, we
can largely decrease the quality of the other. Figure \ref{fig:first_level_graph}(b)
shows the transmission probabilities required to reach $\overline{P}_{S}$
for QRS error correction code for different cases. The solid (dashed)
curves corresponds to the no-multiplexing (multiplexing) cases respectively.
In the multiplexing case we assume that all photons are carrying 4
qubits each. The black curves refer to the situation in which all
photons are traveling in the same channel while the red, yellow and
green curves show the cases in which 5, 10 and 20 qudits are traveling
in the second channel respectively.

Let us now explore this behavior in more detail. For the no-multiplexing
case Fig. \ref{fig:first_level_graph}(b) shows that all curves are
crossing at $p_{t_{2}}\simeq0.94,$ as expected. This is the value
for reaching $\overline{P}_{S}$ when all photons are traveling in
the same channel (solid black line). The advantage of using another
channel, represented by the green (and other color) curves is that
by slightly increasing one channel transmission probability we can
considerably decrease the other. For instance the green curve shows
in the case where channel 2 carries 20 qudits we can by increasing
$p_{t_{2}}$ by approximately $5\%$ to $\sim0.99$ reduce $p_{t_{1}}$
by $\sim13.8\%$ to $0.81.$ This is a significant change in the characteristics
of the channels. This reduction is even more evident for the multiplexing
case represented by the dashed curves in Fig. \ref{fig:first_level_graph}(b).
Here the black dashed curve corresponds to $p_{t_{2}}=0.83$ while
the green dashed curve illustrates the situation with 20 qudits traveling
in the second channel. As an extreme example by improving $p_{t_{2}}$
to $0.99$ $(\sim19.2\%\,\mathrm{increase})$ we can decrease $p_{t_{1}}$
to $0.39$ $(\sim53\%\,\mathrm{lower})$. Alternatively we could increase
$p_{t_{2}}$ to 0.9 $(\sim8.4\%\,\mathrm{increase})$ which in turn
allows us to decrease $p_{t_{1}}$ to $0.757$ $(\sim8.8\%\,\mathrm{lower}).$
Here the aggregate network approach allows to reduce the capacity
of one channel more than the amount the transmission efficiency of
the other channel must be increased. This advantages is enhanced by
applying the quantum multiplexing technique.

Further quantum multiplexing allows to use less photons as well as
less qubits with significant improvements seen as we move to higher
multiplexing degrees \cite{QuM}. However, using photons carrying
an higher number of qubits is not always a convenient strategy to
diminish the number of resources. A practical example will explain
clearly this statement. In the 7-qudit code, each qudit can be encoded
by 3 photons as well as by one photon carrying 3 qubits. Hence multiplexing
the photons by adding an extra qubit, will not be beneficial for the
code because this extra qubit is irrelevant for the encoding of the
qudit and it could be only possibly used to encode a different qudit.
If this happens the loss of this specific photon will destroy both
qudits greatly lowering the success probability. Regardless one can
think of an alternative scenario which we describe in the next Section.

\section{Quantum network aggregation - II }

\textcolor{black}{As highlighted above the loss of a photon that encodes
a qudit will result in the loss of the qudit itself. Therefore when
quantum multiplexing is used it is critical that all the qubits of
a single photon are encoding the same qudit rather than two (or more)
qudits. In this way, the loss of that photon will not affect the loss
of two (or more) qudits. Now by using the aggregating network technique
we have a significant advantage in distributing the qubits of the
same photon over different qudits}. This is our second scenario.
\begin{figure}
\begin{centering}
\includegraphics[scale=0.6]{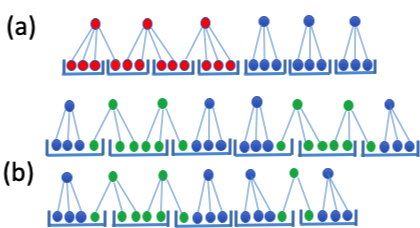}
\par\end{centering}
\caption{\label{fig:sec_lel_schemes}\textcolor{black}{Pictorial representation
of the second aggregate networks scenario in which multiplexed photons
(dots outside the boxes) are encoding different qudits (boxes) for
(a) the 7-qudits QRS code and for (b) the 11-qudits QRS code. In (a)
((b)) the red (green) dots are the photons traveling in the higher
quality channel. Each qudit is encoded with the qubits (dots inside
the boxes) linked to the corresponding photon.}}
\end{figure}
\begin{figure}
\begin{centering}
\includegraphics[scale=0.5]{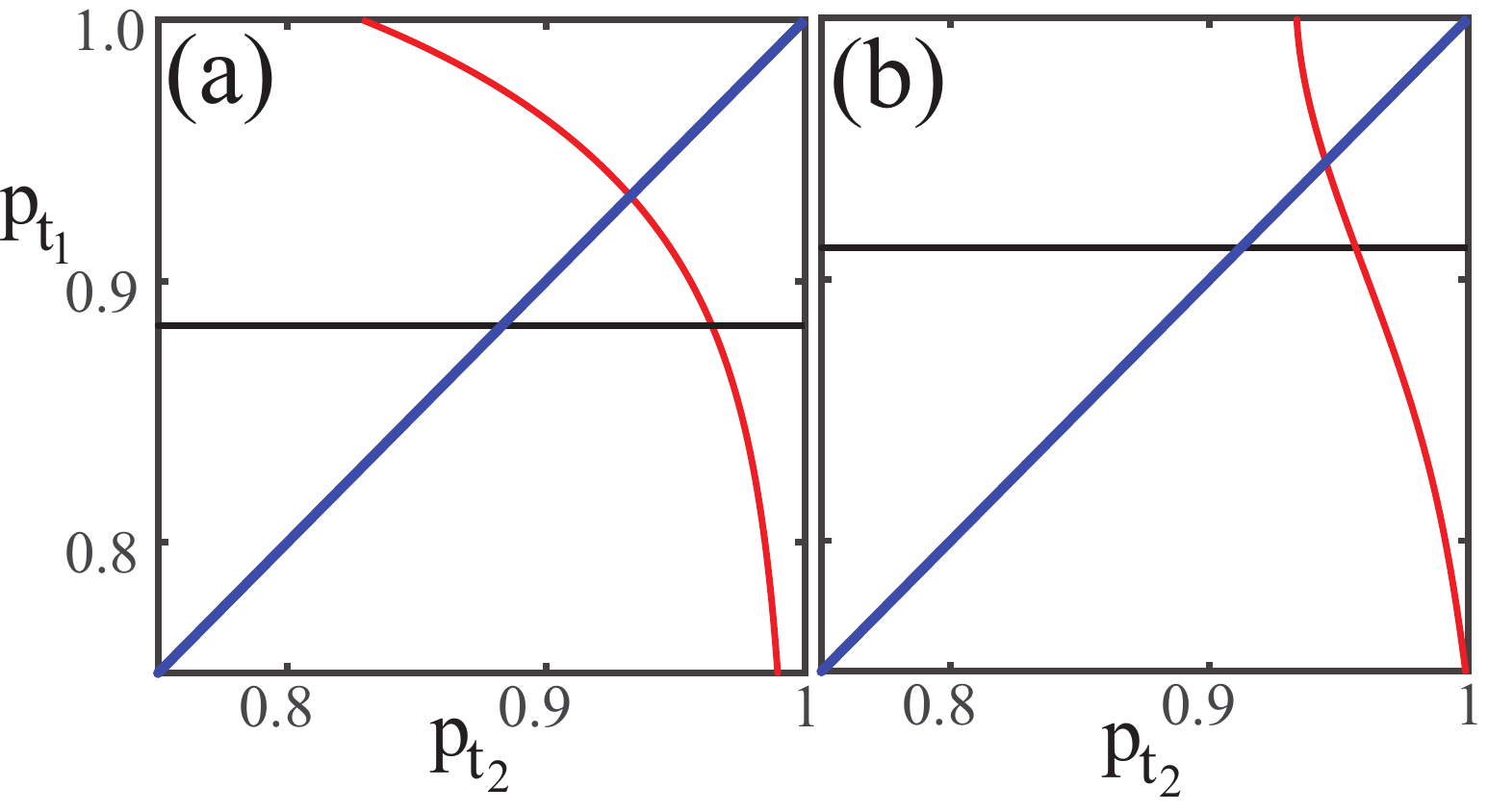}
\par\end{centering}
\caption{\label{fig:7qudits}In (a)((b)), the black curve is the transmission
probability of the 7(11)-qudit QRS code required to reach the threshold
success probability when all photons are carrying 3(2) qubits each.
The red curve gives the nominal values of the transmission probabilities
required to reach the threshold success probability of two channels
for the code represented in Fig. \ref{fig:sec_lel_schemes}(a)((b)),
respectively.}
 
\end{figure}

In this second scenario we assume that multiplexed photons can simultaneously
encode different qudits. In order to compensate for the detrimental
effect arising from loss of such photons they must travel in the higher
quality channel. This scenario is illustrated in Fig. \ref{fig:sec_lel_schemes}(a)((b))
where 6 photons carrying 4 and 3 qubits of information are encoding
seven qudits and in which 15 photons are encoding 11 qudits, respectively.
In Fig. \ref{fig:sec_lel_schemes}(a) we consider the situation where
the red photons are traveling in the better channel whereas in Fig.
\ref{fig:sec_lel_schemes}(b) the green photons are traveling in the
better channel where they encode 2 qudits. Assigning the photons encoding
multiple qudits to the higher transmitivity channel means that the
information has better chance to successfully propagate to the end
of the channel while keeping the success probability of the QRS code
high. We have two success probabilities of interest: first $P_{S_{6}}(2\mathrm{ch})$
representing the situation shown in Fig. \ref{fig:sec_lel_schemes}(a)\textbf{
and second $P_{S_{15}}(2\mathrm{ch})$ representing the situation
shown in Fig. }\ref{fig:sec_lel_schemes}(b). These probabilities
are explicitly given by

\noindent 
\begin{equation}
\begin{array}{c}
P_{S_{6}}(2\mathrm{ch})=p_{t_{1}}^{3}p_{t_{2}}^{3}+{3 \choose {1}}p_{t_{2}}^{2}(1-p_{t_{2}})p_{t_{1}}^{3}+{3 \choose {1}}p_{t_{1}}^{2}(1-p_{t_{1}})p_{t_{2}}^{3}\\
+{3 \choose {1}}p_{t_{2}}^{2}(1-p_{t_{2}}){3 \choose {1}}p_{t_{1}}^{2}(1-p_{t_{1}})+{3 \choose {2}}p_{t_{2}}(1-p_{t_{2}})^{2}p_{t_{1}}^{3}\\
+{3 \choose {2}}p_{t_{2}}^{3}(1-p_{t_{1}})^{2}p_{t_{1}}+p_{t_{2}}^{3}(1-p_{t_{1}})^{3},
\end{array}\label{eq:sec_lev_6ph}
\end{equation}
and 

\begin{equation}
\begin{array}{c}
P_{S_{15}}(2\mathrm{ch})=p_{t_{1}}^{8}p_{t_{2}}^{7}+{7 \choose {1}}p_{t_{2}}^{6}(1-p_{t_{2}})p_{t_{1}}^{8}+{8 \choose {1}}p_{t_{2}}^{7}(1-p_{t_{1}})p_{t_{1}}^{7}\\
+{7 \choose {2}}p_{t_{2}}^{5}(1-p_{t_{2}})p_{t_{1}}^{8}+{8 \choose {2}}p_{t_{2}}^{7}(1-p_{t_{1}})p_{t_{1}}^{6}\\
+{7 \choose {1}}{8 \choose {1}}p_{t_{2}}^{6}(1-p_{t_{2}})(1-p_{t_{1}})p_{t_{1}}^{7}+\\
\left({7 \choose {3}}-20\right)p_{t_{2}}^{4}(1-p_{t_{2}})^{3}p_{t_{1}}^{8}+{8 \choose {3}}p_{t_{2}}^{7}(1-p_{t_{1}})^{3}p_{t_{1}}^{5}\\
+{7 \choose {2}}{8 \choose {1}}p_{t_{2}}^{5}(1-p_{t_{2}})^{2}(1-p_{t_{1}})p_{t_{1}}^{7}\\
+{7 \choose {1}}{8 \choose {2}}p_{t_{2}}^{6}(1-p_{t_{2}})(1-p_{t_{1}})^{2}p_{t_{1}}^{6}\\
+{8 \choose {4}}p_{t_{2}}^{7}(1-p_{t_{1}})^{4}p_{t_{1}}^{4}+{7 \choose {1}}{8 \choose {3}}p_{t_{2}}^{6}(1-p_{t_{2}})(1-p_{t_{1}})^{3}p_{t_{1}}^{5}\\
\left({7 \choose {2}}{8 \choose {2}}-3\left({2 \choose {1}}{2 \choose {1}}{6 \choose {2}}+{2 \choose {1}}{1 \choose {1}}{5 \choose {2}}\right)\right)\\
\times p_{t_{2}}^{5}(1-p_{t_{2}})^{2}(1-p_{t_{1}})^{2}p_{t_{1}}^{6}\\
+\left({7 \choose {3}}{8 \choose {1}}-\left(3\left({2 \choose {2}}{2 \choose {1}}{5 \choose {1}}+{2 \choose {2}}{1 \choose {1}}{4 \choose {2}}\right)+20{8 \choose {1}}\right)\right)\\
\times p_{t_{2}}^{4}(1-p_{t_{2}})^{3}(1-p_{t_{1}})p_{t_{1}}^{7}+{8 \choose {5}}p_{t_{2}}^{7}(1-p_{t_{1}})^{5}p_{t_{1}}^{3}
\end{array}\label{eq:sec_liv_15ph}
\end{equation}
respectively. The exact form of these probabilities depends on the
configuration of distributing the qubits. 

Figures \ref{fig:7qudits}(a) gives a comparison between the transmission
probabilities required to reach $\overline{P}_{S}$ for the 7-qudit
QRS code configured in two different ways. The black curve illustrates
the case in which 7 multiplexed photons ($q=3)$ are traveling through
the same channel while the red curve shows the probabilities for Fig.
\ref{fig:sec_lel_schemes}(a) configuration. Here the transmission
probability of the better channel must be greater than 0.95 if we
want to reduce the transmission probability of the second channel.
This can be a quite big price to pay but the number of photons is
less.

We also observe that the reduction of the number of photons increases
for higher dimensional QRS codes. For instance in the 11-qudit QRS
code we require 22 multiplexed photons each carrying 2 qubits in order
to reach the threshold probability all of them traveling in the same
channel. However we can use the configuration scheme represented in
Fig. \eqref{fig:sec_lel_schemes}(b) to achieve this with only 15
photons by using different degrees of multiplexing. This is illustrated
in Fig. \ref{fig:7qudits}(b) (red curve) and we note that the red
curve crosses the black curve which corresponds to the 11-qudits QRS
code encoded by 22 multiplexed photons. These results show that by
using higher degrees of multiplexing carried by the photons encoding
different qudits, the threshold success probability can be still reached
provided that we use the aggregate network approach. Although not
explicitly shown, we expect a even further reduction of the number
of photons by applying this method to higher dimensional codes. For
instance, for the 43 qudits case, we require 86 photons each of them
carrying 3 qubits traveling in the same channel and only 65 photons
arranged similarly to Fig. \ref{fig:sec_lel_schemes}(b), of which
63 photons carrying 4 qubits and 2 photons carrying 3 qubits. Moreover,
this indicates that, in order to implement this second scenario, the
mixing strate\textcolor{black}{gy, i. e. the procedure of using photons
with different multiplexing degree, is a fundame}ntal approach that
must be used \cite{QuM}. It also highlights a significant advantage
this mixing strategy allows for in quantum network aggregation.

\section{Conclusions}

Future quantum networks will allow the distribution of qubits between
remote users through intermediate nodes connected by multiple paths.
Such routes are likely to have insufficient resources in terms of
number of qubits or channel capacity in order to achieve the successful
transmission of quantum information in all the required situations.
In this work we address such issues by combining our quantum aggregate
network approach with the quantum multiplexing method for the quantum
Reed-Solomon error correction code. We initially illustrate how the
QRS code can be encoded by using photons. This allows us to apply
quantum multiplexing to it, which greatly reduces the number of photons,
qubits and CNOT gates required to reach a given threshold success
probability. We then show however that even for the no multiplexing
case one channel with higher transmission probability can compensate
for the other channel with much lower transmission probability. This
advantage can be further enhanced when quantum multiplexing is used.
In particular for the 43-qudit QRS code, in which each photon has
quantum multiplexing degree equal to 4, and 20 qudits are traveling
in a higher quality channel, the transmission probability of worse
channel can be decrease slightly more than 50\%, while increasing
the transmission probability of only 19.2 \% compared to the case
in which all 43 qudits are traveling in the same channel. 

Further we combine this with the mixed strategy from the quantum multiplexing
scheme, in which the degree of quantum multiplexing may vary for each
photon. In this case the photons traveling in the better channel are
encoding two qudits. Thanks to this configuration the number of photons
encoding the 7-qudit and the 11-qudit QRS codes can be reduced still
reaching the threshold success probability. However in this case the
transmission probability of the better channel must be slightly higher
than the one required for the case in which all photons are traveling
in the same channel. By using this technique, when large dimensional
QRS code are considered, the reduction of photons required is considerably
larger. 

\textcolor{black}{Although not explicitly shown, the same advantages
resulting from the application of this aggregation technique to the
QRS code can be obtained for the quantum parity code and many other
loss based quantum error correction codes. }We believe that aggregate
networks combined with the quantum multiplexing method can be considered
as a valuable approach to alleviate the high costs required for the
implementation of quantum technologies in the near future. 
\begin{acknowledgments}
\textcolor{black}{This project was made possible through the support
of the MEXT KAKENHI Grant-in-Aid for Scientific Research on Innovative
Areas ``Science of Hybrid Quantum Systems\textquotedblright{} Grant
No. 15H05870 and the Japanese MEXT Quantum Leap Flagship Program (MEXT
Q-LEAP), Grant Number JP-MXS0118069605.}
\end{acknowledgments}

\bibliographystyle{apsrev4-1}
\bibliography{bib1}

\begin{thebibliography}{59}%
\makeatletter
\providecommand \@ifxundefined [1]{%
 \@ifx{#1\undefined}
}%
\providecommand \@ifnum [1]{%
 \ifnum #1\expandafter \@firstoftwo
 \else \expandafter \@secondoftwo
 \fi
}%
\providecommand \@ifx [1]{%
 \ifx #1\expandafter \@firstoftwo
 \else \expandafter \@secondoftwo
 \fi
}%
\providecommand \natexlab [1]{#1}%
\providecommand \enquote  [1]{``#1''}%
\providecommand \bibnamefont  [1]{#1}%
\providecommand \bibfnamefont [1]{#1}%
\providecommand \citenamefont [1]{#1}%
\providecommand \href@noop [0]{\@secondoftwo}%
\providecommand \href [0]{\begingroup \@sanitize@url \@href}%
\providecommand \@href[1]{\@@startlink{#1}\@@href}%
\providecommand \@@href[1]{\endgroup#1\@@endlink}%
\providecommand \@sanitize@url [0]{\catcode `\\12\catcode `\$12\catcode
  `\&12\catcode `\#12\catcode `\^12\catcode `\_12\catcode `\%12\relax}%
\providecommand \@@startlink[1]{}%
\providecommand \@@endlink[0]{}%
\providecommand \url  [0]{\begingroup\@sanitize@url \@url }%
\providecommand \@url [1]{\endgroup\@href {#1}{\urlprefix }}%
\providecommand \urlprefix  [0]{URL }%
\providecommand \Eprint [0]{\href }%
\providecommand \doibase [0]{http://dx.doi.org/}%
\providecommand \selectlanguage [0]{\@gobble}%
\providecommand \bibinfo  [0]{\@secondoftwo}%
\providecommand \bibfield  [0]{\@secondoftwo}%
\providecommand \translation [1]{[#1]}%
\providecommand \BibitemOpen [0]{}%
\providecommand \bibitemStop [0]{}%
\providecommand \bibitemNoStop [0]{.\EOS\space}%
\providecommand \EOS [0]{\spacefactor3000\relax}%
\providecommand \BibitemShut  [1]{\csname bibitem#1\endcsname}%
\let\auto@bib@innerbib\@empty
\bibitem [{\citenamefont {Higginbotham}\ \emph {et~al.}(2018)\citenamefont
  {Higginbotham}, \citenamefont {Burns}, \citenamefont {Urmey}, \citenamefont
  {Peterson}, \citenamefont {Kampel}, \citenamefont {Brubaker}, \citenamefont
  {Smith}, \citenamefont {Lehnert},\ and\ \citenamefont {Regal}}]{Qtech}%
  \BibitemOpen
  \bibfield  {author} {\bibinfo {author} {\bibfnamefont {A.~P.}\ \bibnamefont
  {Higginbotham}}, \bibinfo {author} {\bibfnamefont {P.~S.}\ \bibnamefont
  {Burns}}, \bibinfo {author} {\bibfnamefont {M.~D.}\ \bibnamefont {Urmey}},
  \bibinfo {author} {\bibfnamefont {R.~W.}\ \bibnamefont {Peterson}}, \bibinfo
  {author} {\bibfnamefont {N.~S.}\ \bibnamefont {Kampel}}, \bibinfo {author}
  {\bibfnamefont {B.~M.}\ \bibnamefont {Brubaker}}, \bibinfo {author}
  {\bibfnamefont {G.}~\bibnamefont {Smith}}, \bibinfo {author} {\bibfnamefont
  {K.~W.}\ \bibnamefont {Lehnert}}, \ and\ \bibinfo {author} {\bibfnamefont
  {C.~A.}\ \bibnamefont {Regal}},\ }\href@noop {} {\bibfield  {journal}
  {\bibinfo  {journal} {Nature Physics}\ }\textbf {\bibinfo {volume} {14}},\
  \bibinfo {pages} {321} (\bibinfo {year} {2018})}\BibitemShut {NoStop}%
\bibitem [{\citenamefont {Kelly}(2018)}]{Qtech2}%
  \BibitemOpen
  \bibfield  {author} {\bibinfo {author} {\bibfnamefont {J.}~\bibnamefont
  {Kelly}},\ }\href@noop {} {\bibfield  {journal} {\bibinfo  {journal} {Google
  Research Blog}\ } (\bibinfo {year} {2018})}\BibitemShut {NoStop}%
\bibitem [{\citenamefont {Karalic}\ \emph {et~al.}(2020)\citenamefont
  {Karalic}, \citenamefont {Strikalj}, \citenamefont {Masseroni}, \citenamefont
  {Chen}, \citenamefont {Mittag}, \citenamefont {Tschirky}, \citenamefont
  {Wegscheider}, \citenamefont {Ihn}, \citenamefont {Ensslin},\ and\
  \citenamefont {Zilberberg}}]{QDevices1}%
  \BibitemOpen
  \bibfield  {author} {\bibinfo {author} {\bibfnamefont {M.}~\bibnamefont
  {Karalic}}, \bibinfo {author} {\bibfnamefont {A.}~\bibnamefont {Strikalj}},
  \bibinfo {author} {\bibfnamefont {M.}~\bibnamefont {Masseroni}}, \bibinfo
  {author} {\bibfnamefont {W.}~\bibnamefont {Chen}}, \bibinfo {author}
  {\bibfnamefont {C.}~\bibnamefont {Mittag}}, \bibinfo {author} {\bibfnamefont
  {T.}~\bibnamefont {Tschirky}}, \bibinfo {author} {\bibfnamefont
  {W.}~\bibnamefont {Wegscheider}}, \bibinfo {author} {\bibfnamefont
  {T.}~\bibnamefont {Ihn}}, \bibinfo {author} {\bibfnamefont {K.}~\bibnamefont
  {Ensslin}}, \ and\ \bibinfo {author} {\bibfnamefont {O.}~\bibnamefont
  {Zilberberg}},\ }\href@noop {} {\bibfield  {journal} {\bibinfo  {journal}
  {Phys. Rev. X}\ }\textbf {\bibinfo {volume} {10}},\ \bibinfo {pages}
  {0311007} (\bibinfo {year} {2020})}\BibitemShut {NoStop}%
\bibitem [{\citenamefont {Feng}\ \emph {et~al.}(2020)\citenamefont {Feng},
  \citenamefont {Si}, \citenamefont {Min}, \citenamefont {Yuan},\ and\
  \citenamefont {Somekh}}]{QDevices2}%
  \BibitemOpen
  \bibfield  {author} {\bibinfo {author} {\bibfnamefont {F.}~\bibnamefont
  {Feng}}, \bibinfo {author} {\bibfnamefont {G.}~\bibnamefont {Si}}, \bibinfo
  {author} {\bibfnamefont {C.}~\bibnamefont {Min}}, \bibinfo {author}
  {\bibfnamefont {X.}~\bibnamefont {Yuan}}, \ and\ \bibinfo {author}
  {\bibfnamefont {M.}~\bibnamefont {Somekh}},\ }\href@noop {} {\bibfield
  {journal} {\bibinfo  {journal} {Light: Science and Applications}\ }\textbf
  {\bibinfo {volume} {9}} (\bibinfo {year} {2020})}\BibitemShut {NoStop}%
\bibitem [{\citenamefont {Caneva}\ \emph {et~al.}(2020)\citenamefont {Caneva},
  \citenamefont {Hermans}, \citenamefont {Lee}, \citenamefont {Garcia-Fuente},
  \citenamefont {Watanabe}, \citenamefont {Taniguchi}, \citenamefont {Dekker},
  \citenamefont {Ferrer}, \citenamefont {van~der Zant},\ and\ \citenamefont
  {Gehring}}]{QDevices3}%
  \BibitemOpen
  \bibfield  {author} {\bibinfo {author} {\bibfnamefont {S.}~\bibnamefont
  {Caneva}}, \bibinfo {author} {\bibfnamefont {M.}~\bibnamefont {Hermans}},
  \bibinfo {author} {\bibfnamefont {M.}~\bibnamefont {Lee}}, \bibinfo {author}
  {\bibfnamefont {A.}~\bibnamefont {Garcia-Fuente}}, \bibinfo {author}
  {\bibfnamefont {K.}~\bibnamefont {Watanabe}}, \bibinfo {author}
  {\bibfnamefont {T.}~\bibnamefont {Taniguchi}}, \bibinfo {author}
  {\bibfnamefont {C.}~\bibnamefont {Dekker}}, \bibinfo {author} {\bibfnamefont
  {J.}~\bibnamefont {Ferrer}}, \bibinfo {author} {\bibfnamefont {H.~S.~J.}\
  \bibnamefont {van~der Zant}}, \ and\ \bibinfo {author} {\bibfnamefont
  {P.}~\bibnamefont {Gehring}},\ }\href@noop {} {\bibfield  {journal} {\bibinfo
   {journal} {Nano Lett.}\ }\textbf {\bibinfo {volume} {20}},\ \bibinfo {pages}
  {4924} (\bibinfo {year} {2020})}\BibitemShut {NoStop}%
\bibitem [{\citenamefont {Christensen}\ \emph {et~al.}(2020)\citenamefont
  {Christensen}, \citenamefont {Hucul}, \citenamefont {Campbell},\ and\
  \citenamefont {Hudson}}]{QDevices4}%
  \BibitemOpen
  \bibfield  {author} {\bibinfo {author} {\bibfnamefont {J.~E.}\ \bibnamefont
  {Christensen}}, \bibinfo {author} {\bibfnamefont {D.}~\bibnamefont {Hucul}},
  \bibinfo {author} {\bibfnamefont {W.~C.}\ \bibnamefont {Campbell}}, \ and\
  \bibinfo {author} {\bibfnamefont {E.~R.}\ \bibnamefont {Hudson}},\
  }\href@noop {} {\bibfield  {journal} {\bibinfo  {journal} {npj Quantum inf}\
  }\textbf {\bibinfo {volume} {6}},\ \bibinfo {pages} {35} (\bibinfo {year}
  {2020})}\BibitemShut {NoStop}%
\bibitem [{\citenamefont {Anderson}\ \emph {et~al.}(2019)\citenamefont
  {Anderson}, \citenamefont {Bourassa}, \citenamefont {Miao}, \citenamefont
  {Wolfowicz}, \citenamefont {Mintun}, \citenamefont {Crook}, \citenamefont
  {Abe}, \citenamefont {Hassan}, \citenamefont {Son}, \citenamefont {Ohshima},\
  and\ \citenamefont {Awschalom}}]{QDevices5}%
  \BibitemOpen
  \bibfield  {author} {\bibinfo {author} {\bibfnamefont {C.~P.}\ \bibnamefont
  {Anderson}}, \bibinfo {author} {\bibfnamefont {A.}~\bibnamefont {Bourassa}},
  \bibinfo {author} {\bibfnamefont {K.~C.}\ \bibnamefont {Miao}}, \bibinfo
  {author} {\bibfnamefont {G.}~\bibnamefont {Wolfowicz}}, \bibinfo {author}
  {\bibfnamefont {P.~J.}\ \bibnamefont {Mintun}}, \bibinfo {author}
  {\bibfnamefont {A.~L.}\ \bibnamefont {Crook}}, \bibinfo {author}
  {\bibfnamefont {H.}~\bibnamefont {Abe}}, \bibinfo {author} {\bibfnamefont
  {J.~U.}\ \bibnamefont {Hassan}}, \bibinfo {author} {\bibfnamefont {N.~T.}\
  \bibnamefont {Son}}, \bibinfo {author} {\bibfnamefont {T.}~\bibnamefont
  {Ohshima}}, \ and\ \bibinfo {author} {\bibfnamefont {D.~D.}\ \bibnamefont
  {Awschalom}},\ }\href@noop {} {\bibfield  {journal} {\bibinfo  {journal}
  {Science}\ }\textbf {\bibinfo {volume} {366}},\ \bibinfo {pages} {1225}
  (\bibinfo {year} {2019})}\BibitemShut {NoStop}%
\bibitem [{\citenamefont {Nielsen}\ and\ \citenamefont
  {Chuang}(2000)}]{Qcomp1}%
  \BibitemOpen
  \bibfield  {author} {\bibinfo {author} {\bibfnamefont {M.}~\bibnamefont
  {Nielsen}}\ and\ \bibinfo {author} {\bibfnamefont {I.}~\bibnamefont
  {Chuang}},\ }\href@noop {} {\emph {\bibinfo {title} {Quantum Computation and
  Quantum Information}}}\ (\bibinfo  {publisher} {Cambridge University Press,
  Cambridge},\ \bibinfo {year} {2000})\BibitemShut {NoStop}%
\bibitem [{\citenamefont {Bennett}\ and\ \citenamefont
  {DiVincenzo}(2000)}]{Qcomp2}%
  \BibitemOpen
  \bibfield  {author} {\bibinfo {author} {\bibfnamefont {C.}~\bibnamefont
  {Bennett}}\ and\ \bibinfo {author} {\bibfnamefont {D.}~\bibnamefont
  {DiVincenzo}},\ }\href@noop {} {\bibfield  {journal} {\bibinfo  {journal}
  {Nature}\ }\textbf {\bibinfo {volume} {404}},\ \bibinfo {pages} {247}
  (\bibinfo {year} {2000})}\BibitemShut {NoStop}%
\bibitem [{\citenamefont {Raussendorf}\ and\ \citenamefont
  {Briegel}(2001)}]{Quantum_comp2}%
  \BibitemOpen
  \bibfield  {author} {\bibinfo {author} {\bibfnamefont {R.}~\bibnamefont
  {Raussendorf}}\ and\ \bibinfo {author} {\bibfnamefont {H.~J.}\ \bibnamefont
  {Briegel}},\ }\href@noop {} {\bibfield  {journal} {\bibinfo  {journal} {Phys.
  Rev. Lett.}\ }\textbf {\bibinfo {volume} {86}},\ \bibinfo {pages} {5188}
  (\bibinfo {year} {2001})}\BibitemShut {NoStop}%
\bibitem [{\citenamefont {Knill}(2005)}]{Quantum_comp3}%
  \BibitemOpen
  \bibfield  {author} {\bibinfo {author} {\bibfnamefont {E.}~\bibnamefont
  {Knill}},\ }\href@noop {} {\bibfield  {journal} {\bibinfo  {journal}
  {Nature}\ }\textbf {\bibinfo {volume} {434}},\ \bibinfo {pages} {39}
  (\bibinfo {year} {2005})}\BibitemShut {NoStop}%
\bibitem [{\citenamefont {Duan}\ and\ \citenamefont
  {Raussendorf}(2005)}]{Quantum_comp4}%
  \BibitemOpen
  \bibfield  {author} {\bibinfo {author} {\bibfnamefont {L.-M.}\ \bibnamefont
  {Duan}}\ and\ \bibinfo {author} {\bibfnamefont {R.}~\bibnamefont
  {Raussendorf}},\ }\href@noop {} {\bibfield  {journal} {\bibinfo  {journal}
  {Phys. Rev. Lett.}\ }\textbf {\bibinfo {volume} {95}},\ \bibinfo {pages}
  {080503} (\bibinfo {year} {2005})}\BibitemShut {NoStop}%
\bibitem [{\citenamefont {Bennett}\ and\ \citenamefont
  {Brassard}(2014)}]{QComm}%
  \BibitemOpen
  \bibfield  {author} {\bibinfo {author} {\bibfnamefont {C.~H.}\ \bibnamefont
  {Bennett}}\ and\ \bibinfo {author} {\bibfnamefont {G.}~\bibnamefont
  {Brassard}},\ }\href@noop {} {\bibfield  {journal} {\bibinfo  {journal}
  {Theoretical computer science}\ }\textbf {\bibinfo {volume} {560}},\ \bibinfo
  {pages} {7} (\bibinfo {year} {2014})}\BibitemShut {NoStop}%
\bibitem [{\citenamefont {Munro}\ \emph {et~al.}(2015)\citenamefont {Munro},
  \citenamefont {Azuma}, \citenamefont {Tamaki},\ and\ \citenamefont
  {Nemoto}}]{QC_Bill}%
  \BibitemOpen
  \bibfield  {author} {\bibinfo {author} {\bibfnamefont {W.~J.}\ \bibnamefont
  {Munro}}, \bibinfo {author} {\bibfnamefont {K.}~\bibnamefont {Azuma}},
  \bibinfo {author} {\bibfnamefont {K.}~\bibnamefont {Tamaki}}, \ and\ \bibinfo
  {author} {\bibfnamefont {K.}~\bibnamefont {Nemoto}},\ }\href@noop {}
  {\bibfield  {journal} {\bibinfo  {journal} {IEEE Journal of Selected Topics
  in Quantum Electronics}\ }\textbf {\bibinfo {volume} {21}},\ \bibinfo {pages}
  {6400813} (\bibinfo {year} {2015})}\BibitemShut {NoStop}%
\bibitem [{\citenamefont {Sangouard}\ \emph {et~al.}(2011)\citenamefont
  {Sangouard}, \citenamefont {Simon}, \citenamefont {De~Riedmatten},\ and\
  \citenamefont {Gisin}}]{Quantumcomm}%
  \BibitemOpen
  \bibfield  {author} {\bibinfo {author} {\bibfnamefont {N.}~\bibnamefont
  {Sangouard}}, \bibinfo {author} {\bibfnamefont {C.}~\bibnamefont {Simon}},
  \bibinfo {author} {\bibfnamefont {C.}~\bibnamefont {De~Riedmatten}}, \ and\
  \bibinfo {author} {\bibfnamefont {N.}~\bibnamefont {Gisin}},\ }\href@noop {}
  {\bibfield  {journal} {\bibinfo  {journal} {Rev. Mod. Phys.}\ }\textbf
  {\bibinfo {volume} {83}},\ \bibinfo {pages} {33} (\bibinfo {year}
  {2011})}\BibitemShut {NoStop}%
\bibitem [{\citenamefont {Duan}\ \emph {et~al.}(2001)\citenamefont {Duan},
  \citenamefont {Lukin}, \citenamefont {Cirac},\ and\ \citenamefont
  {Zoller}}]{DLCZ_p}%
  \BibitemOpen
  \bibfield  {author} {\bibinfo {author} {\bibfnamefont {L.~M.}\ \bibnamefont
  {Duan}}, \bibinfo {author} {\bibfnamefont {M.~D.}\ \bibnamefont {Lukin}},
  \bibinfo {author} {\bibfnamefont {I.}~\bibnamefont {Cirac}}, \ and\ \bibinfo
  {author} {\bibfnamefont {P.}~\bibnamefont {Zoller}},\ }\href@noop {}
  {\bibfield  {journal} {\bibinfo  {journal} {Nature}\ }\textbf {\bibinfo
  {volume} {414}},\ \bibinfo {pages} {413} (\bibinfo {year}
  {2001})}\BibitemShut {NoStop}%
\bibitem [{\citenamefont {Lo}(1999)}]{QKD02}%
  \BibitemOpen
  \bibfield  {author} {\bibinfo {author} {\bibfnamefont {H.}~\bibnamefont
  {Lo}},\ }\href@noop {} {\bibfield  {journal} {\bibinfo  {journal} {Science}\
  }\textbf {\bibinfo {volume} {283}},\ \bibinfo {pages} {2050} (\bibinfo {year}
  {1999})}\BibitemShut {NoStop}%
\bibitem [{\citenamefont {Hwang}(2003)}]{QKD03}%
  \BibitemOpen
  \bibfield  {author} {\bibinfo {author} {\bibfnamefont {W.-Y.}\ \bibnamefont
  {Hwang}},\ }\href@noop {} {\bibfield  {journal} {\bibinfo  {journal} {Phys.
  Rev. Lett.}\ }\textbf {\bibinfo {volume} {91}},\ \bibinfo {pages} {057901}
  (\bibinfo {year} {2003})}\BibitemShut {NoStop}%
\bibitem [{\citenamefont {Ekert}(1991)}]{QKD1}%
  \BibitemOpen
  \bibfield  {author} {\bibinfo {author} {\bibfnamefont {A.~K.}\ \bibnamefont
  {Ekert}},\ }\href@noop {} {\bibfield  {journal} {\bibinfo  {journal} {Phys.
  Rev. Lett.}\ }\textbf {\bibinfo {volume} {67}},\ \bibinfo {pages} {661}
  (\bibinfo {year} {1991})}\BibitemShut {NoStop}%
\bibitem [{\citenamefont {Qiu}(2014)}]{QKD2000}%
  \BibitemOpen
  \bibfield  {author} {\bibinfo {author} {\bibfnamefont {J.}~\bibnamefont
  {Qiu}},\ }\href@noop {} {\bibfield  {journal} {\bibinfo  {journal} {Nature}\
  }\textbf {\bibinfo {volume} {508}},\ \bibinfo {pages} {441} (\bibinfo {year}
  {2014})}\BibitemShut {NoStop}%
\bibitem [{\citenamefont {Bunandar}\ \emph {et~al.}(2018)\citenamefont
  {Bunandar}, \citenamefont {Lentine}, \citenamefont {Lee}, \citenamefont
  {Cai}, \citenamefont {Long}, \citenamefont {Boynton}, \citenamefont
  {Martinez}, \citenamefont {DeRose}, \citenamefont {Chen}, \citenamefont
  {Grein}, \citenamefont {Trotter}, \citenamefont {Starbuck}, \citenamefont
  {Pomerene}, \citenamefont {Hamilton}, \citenamefont {Wong}, \citenamefont
  {Camacho}, \citenamefont {Davids}, \citenamefont {Urayama},\ and\
  \citenamefont {Englund}}]{QKD_transmission}%
  \BibitemOpen
  \bibfield  {author} {\bibinfo {author} {\bibfnamefont {D.}~\bibnamefont
  {Bunandar}}, \bibinfo {author} {\bibfnamefont {A.}~\bibnamefont {Lentine}},
  \bibinfo {author} {\bibfnamefont {C.}~\bibnamefont {Lee}}, \bibinfo {author}
  {\bibfnamefont {H.}~\bibnamefont {Cai}}, \bibinfo {author} {\bibfnamefont
  {C.~M.}\ \bibnamefont {Long}}, \bibinfo {author} {\bibfnamefont
  {N.}~\bibnamefont {Boynton}}, \bibinfo {author} {\bibfnamefont
  {N.}~\bibnamefont {Martinez}}, \bibinfo {author} {\bibfnamefont
  {C.}~\bibnamefont {DeRose}}, \bibinfo {author} {\bibfnamefont
  {C.}~\bibnamefont {Chen}}, \bibinfo {author} {\bibfnamefont {M.}~\bibnamefont
  {Grein}}, \bibinfo {author} {\bibfnamefont {D.}~\bibnamefont {Trotter}},
  \bibinfo {author} {\bibfnamefont {A.}~\bibnamefont {Starbuck}}, \bibinfo
  {author} {\bibfnamefont {A.}~\bibnamefont {Pomerene}}, \bibinfo {author}
  {\bibfnamefont {S.}~\bibnamefont {Hamilton}}, \bibinfo {author}
  {\bibfnamefont {F.~N.~C.}\ \bibnamefont {Wong}}, \bibinfo {author}
  {\bibfnamefont {R.}~\bibnamefont {Camacho}}, \bibinfo {author} {\bibfnamefont
  {P.}~\bibnamefont {Davids}}, \bibinfo {author} {\bibfnamefont
  {J.}~\bibnamefont {Urayama}}, \ and\ \bibinfo {author} {\bibfnamefont
  {D.}~\bibnamefont {Englund}},\ }\href@noop {} {\bibfield  {journal} {\bibinfo
   {journal} {Phys. Rev. X}\ }\textbf {\bibinfo {volume} {8}},\ \bibinfo
  {pages} {12} (\bibinfo {year} {2018})}\BibitemShut {NoStop}%
\bibitem [{\citenamefont {Toth}(2012)}]{Qmetrology1}%
  \BibitemOpen
  \bibfield  {author} {\bibinfo {author} {\bibfnamefont {G.}~\bibnamefont
  {Toth}},\ }\href@noop {} {\bibfield  {journal} {\bibinfo  {journal} {Phys.
  Rev. A}\ }\textbf {\bibinfo {volume} {85}},\ \bibinfo {pages} {022322}
  (\bibinfo {year} {2012})}\BibitemShut {NoStop}%
\bibitem [{\citenamefont {Guo-Yong}\ and\ \citenamefont
  {Guang-Can}(2013)}]{Qmetrology2}%
  \BibitemOpen
  \bibfield  {author} {\bibinfo {author} {\bibfnamefont {X.}~\bibnamefont
  {Guo-Yong}}\ and\ \bibinfo {author} {\bibfnamefont {G.}~\bibnamefont
  {Guang-Can}},\ }\href@noop {} {\bibfield  {journal} {\bibinfo  {journal}
  {Chinese Physics B}\ }\textbf {\bibinfo {volume} {22}} (\bibinfo {year}
  {2013})}\BibitemShut {NoStop}%
\bibitem [{\citenamefont {Dogen}\ \emph {et~al.}(2017)\citenamefont {Dogen},
  \citenamefont {Reinhard},\ and\ \citenamefont {Cappellaro}}]{QSensing}%
  \BibitemOpen
  \bibfield  {author} {\bibinfo {author} {\bibfnamefont {C.~L.}\ \bibnamefont
  {Dogen}}, \bibinfo {author} {\bibfnamefont {F.}~\bibnamefont {Reinhard}}, \
  and\ \bibinfo {author} {\bibfnamefont {P.}~\bibnamefont {Cappellaro}},\
  }\href@noop {} {\bibfield  {journal} {\bibinfo  {journal} {Rev. Mod. Phys.}\
  }\textbf {\bibinfo {volume} {89}},\ \bibinfo {pages} {035002} (\bibinfo
  {year} {2017})}\BibitemShut {NoStop}%
\bibitem [{\citenamefont {Lugiato}\ \emph {et~al.}(2002)\citenamefont
  {Lugiato}, \citenamefont {Gatti},\ and\ \citenamefont
  {Brambilla}}]{QImaging3}%
  \BibitemOpen
  \bibfield  {author} {\bibinfo {author} {\bibfnamefont {L.~A.}\ \bibnamefont
  {Lugiato}}, \bibinfo {author} {\bibfnamefont {A.}~\bibnamefont {Gatti}}, \
  and\ \bibinfo {author} {\bibfnamefont {E.}~\bibnamefont {Brambilla}},\
  }\href@noop {} {\bibfield  {journal} {\bibinfo  {journal} {J. Opt. B}\
  }\textbf {\bibinfo {volume} {4}},\ \bibinfo {pages} {176} (\bibinfo {year}
  {2002})}\BibitemShut {NoStop}%
\bibitem [{\citenamefont {Simon}\ \emph {et~al.}(2014)\citenamefont {Simon},
  \citenamefont {Jaeger},\ and\ \citenamefont {Sergienko}}]{QImaging2}%
  \BibitemOpen
  \bibfield  {author} {\bibinfo {author} {\bibfnamefont {D.~S.}\ \bibnamefont
  {Simon}}, \bibinfo {author} {\bibfnamefont {G.}~\bibnamefont {Jaeger}}, \
  and\ \bibinfo {author} {\bibfnamefont {A.~V.}\ \bibnamefont {Sergienko}},\
  }\href@noop {} {\bibfield  {journal} {\bibinfo  {journal} {Int. J. Quantum
  Inform.}\ }\textbf {\bibinfo {volume} {12}},\ \bibinfo {pages} {1430004}
  (\bibinfo {year} {2014})}\BibitemShut {NoStop}%
\bibitem [{\citenamefont {Arute}\ \emph {et~al.}(2019)\citenamefont {Arute},
  \citenamefont {Arya}, \citenamefont {Babbush}, \citenamefont {Bacon},
  \citenamefont {Bardin}, \citenamefont {Barends}, \citenamefont {Biswas},
  \citenamefont {Boixo}, \citenamefont {Brandao}, \citenamefont {Buell},
  \citenamefont {Burkett}, \citenamefont {Chen}, \citenamefont {Chen},
  \citenamefont {Chiaro}, \citenamefont {Collins}, \citenamefont {Courtney},
  \citenamefont {Dunsworth}, \citenamefont {Farhi}, \citenamefont {Foxen},
  \citenamefont {Fowler}, \citenamefont {Gidney}, \citenamefont {Giustina},
  \citenamefont {Graff}, \citenamefont {Guerin}, \citenamefont {Habegger},
  \citenamefont {Harrigan}, \citenamefont {Hartmann}, \citenamefont {Ho},
  \citenamefont {Hoffmann}, \citenamefont {Huang}, \citenamefont {Humble},
  \citenamefont {I.}, \citenamefont {Jeffrey}, \citenamefont {Jiang},
  \citenamefont {Kafri}, \citenamefont {Kechedzhi}, \citenamefont {Kelly},
  \citenamefont {Klimov}, \citenamefont {Knysh}, \citenamefont {Korotkov},
  \citenamefont {Kostritsa}, \citenamefont {Landhuis}, \citenamefont
  {Lindmark}, \citenamefont {Lucero}, \citenamefont {Lyakh}, \citenamefont
  {Mandrà}, \citenamefont {McClean}, \citenamefont {McEwen}, \citenamefont
  {Megrant}, \citenamefont {Mi}, \citenamefont {Michielsen}, \citenamefont
  {Mohseni}, \citenamefont {Mutus}, \citenamefont {Naaman}, \citenamefont
  {Neeley}, \citenamefont {Neill}, \citenamefont {Niu}, \citenamefont {Ostby},
  \citenamefont {Petukhov}, \citenamefont {Platt}, \citenamefont {Quintana},
  \citenamefont {Rieffel}, \citenamefont {Roushan}, \citenamefont {Rubin},
  \citenamefont {Sank}, \citenamefont {Satzinger}, \citenamefont {Smelyanskiy},
  \citenamefont {Sung}, \citenamefont {Trevithick}, \citenamefont
  {Vainsencher}, \citenamefont {Villalonga}, \citenamefont {White},
  \citenamefont {Yao}, \citenamefont {Yeh}, \citenamefont {Zalcman},
  \citenamefont {Neven},\ and\ \citenamefont {Martinis}}]{QC_google}%
  \BibitemOpen
  \bibfield  {author} {\bibinfo {author} {\bibfnamefont {F.}~\bibnamefont
  {Arute}}, \bibinfo {author} {\bibfnamefont {K.}~\bibnamefont {Arya}},
  \bibinfo {author} {\bibfnamefont {R.}~\bibnamefont {Babbush}}, \bibinfo
  {author} {\bibfnamefont {D.}~\bibnamefont {Bacon}}, \bibinfo {author}
  {\bibfnamefont {J.~C.}\ \bibnamefont {Bardin}}, \bibinfo {author}
  {\bibfnamefont {R.}~\bibnamefont {Barends}}, \bibinfo {author} {\bibfnamefont
  {R.}~\bibnamefont {Biswas}}, \bibinfo {author} {\bibfnamefont
  {S.}~\bibnamefont {Boixo}}, \bibinfo {author} {\bibfnamefont {F.~G. S.~L.}\
  \bibnamefont {Brandao}}, \bibinfo {author} {\bibfnamefont {D.~A.}\
  \bibnamefont {Buell}}, \bibinfo {author} {\bibfnamefont {B.}~\bibnamefont
  {Burkett}}, \bibinfo {author} {\bibfnamefont {Y.}~\bibnamefont {Chen}},
  \bibinfo {author} {\bibfnamefont {Z.}~\bibnamefont {Chen}}, \bibinfo {author}
  {\bibfnamefont {B.}~\bibnamefont {Chiaro}}, \bibinfo {author} {\bibfnamefont
  {R.}~\bibnamefont {Collins}}, \bibinfo {author} {\bibfnamefont
  {W.}~\bibnamefont {Courtney}}, \bibinfo {author} {\bibfnamefont
  {A.}~\bibnamefont {Dunsworth}}, \bibinfo {author} {\bibfnamefont
  {E.}~\bibnamefont {Farhi}}, \bibinfo {author} {\bibfnamefont
  {B.}~\bibnamefont {Foxen}}, \bibinfo {author} {\bibfnamefont
  {A.}~\bibnamefont {Fowler}}, \bibinfo {author} {\bibfnamefont
  {C.}~\bibnamefont {Gidney}}, \bibinfo {author} {\bibfnamefont
  {M.}~\bibnamefont {Giustina}}, \bibinfo {author} {\bibfnamefont
  {R.}~\bibnamefont {Graff}}, \bibinfo {author} {\bibfnamefont
  {K.}~\bibnamefont {Guerin}}, \bibinfo {author} {\bibfnamefont
  {S.}~\bibnamefont {Habegger}}, \bibinfo {author} {\bibfnamefont {M.~P.}\
  \bibnamefont {Harrigan}}, \bibinfo {author} {\bibfnamefont {M.~J.}\
  \bibnamefont {Hartmann}}, \bibinfo {author} {\bibfnamefont {A.}~\bibnamefont
  {Ho}}, \bibinfo {author} {\bibfnamefont {M.}~\bibnamefont {Hoffmann}},
  \bibinfo {author} {\bibfnamefont {T.}~\bibnamefont {Huang}}, \bibinfo
  {author} {\bibfnamefont {T.~S.}\ \bibnamefont {Humble}}, \bibinfo {author}
  {\bibfnamefont {S.~V.}\ \bibnamefont {I.}}, \bibinfo {author} {\bibfnamefont
  {E.}~\bibnamefont {Jeffrey}}, \bibinfo {author} {\bibfnamefont
  {Z.}~\bibnamefont {Jiang}}, \bibinfo {author} {\bibfnamefont
  {D.}~\bibnamefont {Kafri}}, \bibinfo {author} {\bibfnamefont
  {K.}~\bibnamefont {Kechedzhi}}, \bibinfo {author} {\bibfnamefont
  {J.}~\bibnamefont {Kelly}}, \bibinfo {author} {\bibfnamefont {P.~V.}\
  \bibnamefont {Klimov}}, \bibinfo {author} {\bibfnamefont {S.}~\bibnamefont
  {Knysh}}, \bibinfo {author} {\bibfnamefont {A.}~\bibnamefont {Korotkov}},
  \bibinfo {author} {\bibfnamefont {F.}~\bibnamefont {Kostritsa}}, \bibinfo
  {author} {\bibfnamefont {D.}~\bibnamefont {Landhuis}}, \bibinfo {author}
  {\bibfnamefont {M.}~\bibnamefont {Lindmark}}, \bibinfo {author}
  {\bibfnamefont {E.}~\bibnamefont {Lucero}}, \bibinfo {author} {\bibfnamefont
  {D.}~\bibnamefont {Lyakh}}, \bibinfo {author} {\bibfnamefont
  {S.}~\bibnamefont {Mandrà}}, \bibinfo {author} {\bibfnamefont {J.~R.}\
  \bibnamefont {McClean}}, \bibinfo {author} {\bibfnamefont {M.}~\bibnamefont
  {McEwen}}, \bibinfo {author} {\bibfnamefont {A.}~\bibnamefont {Megrant}},
  \bibinfo {author} {\bibfnamefont {X.}~\bibnamefont {Mi}}, \bibinfo {author}
  {\bibfnamefont {K.}~\bibnamefont {Michielsen}}, \bibinfo {author}
  {\bibfnamefont {M.}~\bibnamefont {Mohseni}}, \bibinfo {author} {\bibfnamefont
  {J.}~\bibnamefont {Mutus}}, \bibinfo {author} {\bibfnamefont
  {O.}~\bibnamefont {Naaman}}, \bibinfo {author} {\bibfnamefont
  {M.}~\bibnamefont {Neeley}}, \bibinfo {author} {\bibfnamefont
  {C.}~\bibnamefont {Neill}}, \bibinfo {author} {\bibfnamefont {M.~Y.}\
  \bibnamefont {Niu}}, \bibinfo {author} {\bibfnamefont {E.}~\bibnamefont
  {Ostby}}, \bibinfo {author} {\bibfnamefont {A.}~\bibnamefont {Petukhov}},
  \bibinfo {author} {\bibfnamefont {J.~C.}\ \bibnamefont {Platt}}, \bibinfo
  {author} {\bibfnamefont {C.}~\bibnamefont {Quintana}}, \bibinfo {author}
  {\bibfnamefont {E.~G.}\ \bibnamefont {Rieffel}}, \bibinfo {author}
  {\bibfnamefont {P.}~\bibnamefont {Roushan}}, \bibinfo {author} {\bibfnamefont
  {N.~C.}\ \bibnamefont {Rubin}}, \bibinfo {author} {\bibfnamefont
  {D.}~\bibnamefont {Sank}}, \bibinfo {author} {\bibfnamefont {K.~J.}\
  \bibnamefont {Satzinger}}, \bibinfo {author} {\bibfnamefont {V.}~\bibnamefont
  {Smelyanskiy}}, \bibinfo {author} {\bibfnamefont {K.~J.}\ \bibnamefont
  {Sung}}, \bibinfo {author} {\bibfnamefont {M.~D.}\ \bibnamefont
  {Trevithick}}, \bibinfo {author} {\bibfnamefont {A.}~\bibnamefont
  {Vainsencher}}, \bibinfo {author} {\bibfnamefont {B.}~\bibnamefont
  {Villalonga}}, \bibinfo {author} {\bibfnamefont {T.}~\bibnamefont {White}},
  \bibinfo {author} {\bibfnamefont {Z.~J.}\ \bibnamefont {Yao}}, \bibinfo
  {author} {\bibfnamefont {P.}~\bibnamefont {Yeh}}, \bibinfo {author}
  {\bibfnamefont {A.}~\bibnamefont {Zalcman}}, \bibinfo {author} {\bibfnamefont
  {H.}~\bibnamefont {Neven}}, \ and\ \bibinfo {author} {\bibfnamefont {J.~M.}\
  \bibnamefont {Martinis}},\ }\href@noop {} {\bibfield  {journal} {\bibinfo
  {journal} {Nature}\ }\textbf {\bibinfo {volume} {574}},\ \bibinfo {pages}
  {505} (\bibinfo {year} {2019})}\BibitemShut {NoStop}%
\bibitem [{\citenamefont {Liao}\ \emph {et~al.}(2017)\citenamefont {Liao},
  \citenamefont {Cai}, \citenamefont {Liu}, \citenamefont {Zhang},
  \citenamefont {Li}, \citenamefont {Ren}, \citenamefont {Yin}, \citenamefont
  {Shen}, \citenamefont {Cao}, \citenamefont {Li}, \citenamefont {Li},
  \citenamefont {Chen}, \citenamefont {Sun}, \citenamefont {Jia}, \citenamefont
  {Wu}, \citenamefont {Jiang}, \citenamefont {Wang}, \citenamefont {Huang},
  \citenamefont {Wang}, \citenamefont {Zhou}, \citenamefont {Deng},
  \citenamefont {Xi}, \citenamefont {Ma}, \citenamefont {Hu}, \citenamefont
  {Zhang}, \citenamefont {Chen}, \citenamefont {Liu}, \citenamefont {Wang},
  \citenamefont {Zhu}, \citenamefont {Lu}, \citenamefont {Shu}, \citenamefont
  {Peng}, \citenamefont {Wang},\ and\ \citenamefont {Pan}}]{satelliteQKD}%
  \BibitemOpen
  \bibfield  {author} {\bibinfo {author} {\bibfnamefont {S.-K.}\ \bibnamefont
  {Liao}}, \bibinfo {author} {\bibfnamefont {W.-Q.}\ \bibnamefont {Cai}},
  \bibinfo {author} {\bibfnamefont {W.-Y.}\ \bibnamefont {Liu}}, \bibinfo
  {author} {\bibfnamefont {L.}~\bibnamefont {Zhang}}, \bibinfo {author}
  {\bibfnamefont {Y.}~\bibnamefont {Li}}, \bibinfo {author} {\bibfnamefont
  {J.-G.}\ \bibnamefont {Ren}}, \bibinfo {author} {\bibfnamefont
  {J.}~\bibnamefont {Yin}}, \bibinfo {author} {\bibfnamefont {Q.}~\bibnamefont
  {Shen}}, \bibinfo {author} {\bibfnamefont {Y.}~\bibnamefont {Cao}}, \bibinfo
  {author} {\bibfnamefont {Z.-P.}\ \bibnamefont {Li}}, \bibinfo {author}
  {\bibfnamefont {F.-Z.}\ \bibnamefont {Li}}, \bibinfo {author} {\bibfnamefont
  {X.-W.}\ \bibnamefont {Chen}}, \bibinfo {author} {\bibfnamefont {L.-H.}\
  \bibnamefont {Sun}}, \bibinfo {author} {\bibfnamefont {J.-J.}\ \bibnamefont
  {Jia}}, \bibinfo {author} {\bibfnamefont {J.-C.}\ \bibnamefont {Wu}},
  \bibinfo {author} {\bibfnamefont {X.-J.}\ \bibnamefont {Jiang}}, \bibinfo
  {author} {\bibfnamefont {J.-F.}\ \bibnamefont {Wang}}, \bibinfo {author}
  {\bibfnamefont {Y.-M.}\ \bibnamefont {Huang}}, \bibinfo {author}
  {\bibfnamefont {Q.}~\bibnamefont {Wang}}, \bibinfo {author} {\bibfnamefont
  {Y.-L.}\ \bibnamefont {Zhou}}, \bibinfo {author} {\bibfnamefont
  {L.}~\bibnamefont {Deng}}, \bibinfo {author} {\bibfnamefont {T.}~\bibnamefont
  {Xi}}, \bibinfo {author} {\bibfnamefont {L.}~\bibnamefont {Ma}}, \bibinfo
  {author} {\bibfnamefont {T.}~\bibnamefont {Hu}}, \bibinfo {author}
  {\bibfnamefont {Q.}~\bibnamefont {Zhang}}, \bibinfo {author} {\bibfnamefont
  {Y.-A.}\ \bibnamefont {Chen}}, \bibinfo {author} {\bibfnamefont {N.-L.}\
  \bibnamefont {Liu}}, \bibinfo {author} {\bibfnamefont {X.-B.}\ \bibnamefont
  {Wang}}, \bibinfo {author} {\bibfnamefont {Z.-C.}\ \bibnamefont {Zhu}},
  \bibinfo {author} {\bibfnamefont {C.-Y.}\ \bibnamefont {Lu}}, \bibinfo
  {author} {\bibfnamefont {R.}~\bibnamefont {Shu}}, \bibinfo {author}
  {\bibfnamefont {C.-Z.}\ \bibnamefont {Peng}}, \bibinfo {author}
  {\bibfnamefont {J.-Y.}\ \bibnamefont {Wang}}, \ and\ \bibinfo {author}
  {\bibfnamefont {J.-W.}\ \bibnamefont {Pan}},\ }\href@noop {} {\bibfield
  {journal} {\bibinfo  {journal} {Nature}\ }\textbf {\bibinfo {volume} {549}},\
  \bibinfo {pages} {43} (\bibinfo {year} {2017})}\BibitemShut {NoStop}%
\bibitem [{\citenamefont {Kimble}(2008)}]{Q_internet}%
  \BibitemOpen
  \bibfield  {author} {\bibinfo {author} {\bibfnamefont {H.~J.}\ \bibnamefont
  {Kimble}},\ }\href@noop {} {\bibfield  {journal} {\bibinfo  {journal}
  {Nature}\ }\textbf {\bibinfo {volume} {453}},\ \bibinfo {pages} {1023}
  (\bibinfo {year} {2008})}\BibitemShut {NoStop}%
\bibitem [{\citenamefont {Munro}\ \emph {et~al.}(2012)\citenamefont {Munro},
  \citenamefont {Stephens}, \citenamefont {Devitt}, \citenamefont {Harrison},\
  and\ \citenamefont {Nemoto}}]{Bill_nature_phot}%
  \BibitemOpen
  \bibfield  {author} {\bibinfo {author} {\bibfnamefont {W.~J.}\ \bibnamefont
  {Munro}}, \bibinfo {author} {\bibfnamefont {A.~M.}\ \bibnamefont {Stephens}},
  \bibinfo {author} {\bibfnamefont {S.~J.}\ \bibnamefont {Devitt}}, \bibinfo
  {author} {\bibfnamefont {K.~A.}\ \bibnamefont {Harrison}}, \ and\ \bibinfo
  {author} {\bibfnamefont {K.}~\bibnamefont {Nemoto}},\ }\href@noop {}
  {\bibfield  {journal} {\bibinfo  {journal} {Nature Photonics}\ }\textbf
  {\bibinfo {volume} {6}},\ \bibinfo {pages} {777} (\bibinfo {year}
  {2012})}\BibitemShut {NoStop}%
\bibitem [{\citenamefont {Bennett}\ \emph {et~al.}(1996)\citenamefont
  {Bennett}, \citenamefont {Brassard}, \citenamefont {Popesco}, \citenamefont
  {Schumacher}, \citenamefont {Smolin},\ and\ \citenamefont
  {Wootters}}]{tel_1}%
  \BibitemOpen
  \bibfield  {author} {\bibinfo {author} {\bibfnamefont {C.}~\bibnamefont
  {Bennett}}, \bibinfo {author} {\bibfnamefont {G.}~\bibnamefont {Brassard}},
  \bibinfo {author} {\bibfnamefont {S.}~\bibnamefont {Popesco}}, \bibinfo
  {author} {\bibfnamefont {B.}~\bibnamefont {Schumacher}}, \bibinfo {author}
  {\bibfnamefont {J.}~\bibnamefont {Smolin}}, \ and\ \bibinfo {author}
  {\bibfnamefont {W.}~\bibnamefont {Wootters}},\ }\href@noop {} {\bibfield
  {journal} {\bibinfo  {journal} {Phys. Rev. Lett.}\ }\textbf {\bibinfo
  {volume} {76}},\ \bibinfo {pages} {722} (\bibinfo {year} {1996})}\BibitemShut
  {NoStop}%
\bibitem [{\citenamefont {Deng}\ \emph {et~al.}(2005)\citenamefont {Deng},
  \citenamefont {Li}, \citenamefont {Li}, \citenamefont {Zhou},\ and\
  \citenamefont {Wang}}]{tel_2}%
  \BibitemOpen
  \bibfield  {author} {\bibinfo {author} {\bibfnamefont {F.}~\bibnamefont
  {Deng}}, \bibinfo {author} {\bibfnamefont {C.}~\bibnamefont {Li}}, \bibinfo
  {author} {\bibfnamefont {Y.}~\bibnamefont {Li}}, \bibinfo {author}
  {\bibfnamefont {H.}~\bibnamefont {Zhou}}, \ and\ \bibinfo {author}
  {\bibfnamefont {Y.}~\bibnamefont {Wang}},\ }\href@noop {} {\bibfield
  {journal} {\bibinfo  {journal} {Phys. Rev. A}\ }\textbf {\bibinfo {volume}
  {72}},\ \bibinfo {pages} {022338} (\bibinfo {year} {2005})}\BibitemShut
  {NoStop}%
\bibitem [{\citenamefont {Barrett}\ \emph {et~al.}(2004)\citenamefont
  {Barrett}, \citenamefont {Chiaverini}, \citenamefont {Schaetz}, \citenamefont
  {Britton}, \citenamefont {Itano}, \citenamefont {Jost}, \citenamefont
  {Knill}, \citenamefont {Langer}, \citenamefont {Leibfried}, \citenamefont
  {Ozeri},\ and\ \citenamefont {Wineland}}]{Qtel2}%
  \BibitemOpen
  \bibfield  {author} {\bibinfo {author} {\bibfnamefont {M.~D.}\ \bibnamefont
  {Barrett}}, \bibinfo {author} {\bibfnamefont {J.}~\bibnamefont {Chiaverini}},
  \bibinfo {author} {\bibfnamefont {T.}~\bibnamefont {Schaetz}}, \bibinfo
  {author} {\bibfnamefont {J.}~\bibnamefont {Britton}}, \bibinfo {author}
  {\bibfnamefont {W.~M.}\ \bibnamefont {Itano}}, \bibinfo {author}
  {\bibfnamefont {J.~D.}\ \bibnamefont {Jost}}, \bibinfo {author}
  {\bibfnamefont {E.}~\bibnamefont {Knill}}, \bibinfo {author} {\bibfnamefont
  {C.}~\bibnamefont {Langer}}, \bibinfo {author} {\bibfnamefont
  {D.}~\bibnamefont {Leibfried}}, \bibinfo {author} {\bibfnamefont
  {R.}~\bibnamefont {Ozeri}}, \ and\ \bibinfo {author} {\bibfnamefont {D.~J.}\
  \bibnamefont {Wineland}},\ }\href@noop {} {\bibfield  {journal} {\bibinfo
  {journal} {Nature}\ }\textbf {\bibinfo {volume} {429}},\ \bibinfo {pages}
  {737} (\bibinfo {year} {2004})}\BibitemShut {NoStop}%
\bibitem [{\citenamefont {Ma}\ \emph {et~al.}(2012)\citenamefont {Ma},
  \citenamefont {Herbst}, \citenamefont {Scheidl}, \citenamefont {Wang},
  \citenamefont {Kropatschenk}, \citenamefont {Naylor}, \citenamefont
  {Wittmann}, \citenamefont {Mech}, \citenamefont {Kofler}, \citenamefont
  {Anisimova}, \citenamefont {Marakov}, \citenamefont {Jennewein},
  \citenamefont {Ursin},\ and\ \citenamefont {Zeilinger}}]{Qtel3}%
  \BibitemOpen
  \bibfield  {author} {\bibinfo {author} {\bibfnamefont {X.-S.}\ \bibnamefont
  {Ma}}, \bibinfo {author} {\bibfnamefont {T.}~\bibnamefont {Herbst}}, \bibinfo
  {author} {\bibfnamefont {T.}~\bibnamefont {Scheidl}}, \bibinfo {author}
  {\bibfnamefont {D.}~\bibnamefont {Wang}}, \bibinfo {author} {\bibfnamefont
  {S.}~\bibnamefont {Kropatschenk}}, \bibinfo {author} {\bibfnamefont
  {W.}~\bibnamefont {Naylor}}, \bibinfo {author} {\bibfnamefont
  {B.}~\bibnamefont {Wittmann}}, \bibinfo {author} {\bibfnamefont
  {A.}~\bibnamefont {Mech}}, \bibinfo {author} {\bibfnamefont {J.}~\bibnamefont
  {Kofler}}, \bibinfo {author} {\bibfnamefont {E.}~\bibnamefont {Anisimova}},
  \bibinfo {author} {\bibfnamefont {V.}~\bibnamefont {Marakov}}, \bibinfo
  {author} {\bibfnamefont {T.}~\bibnamefont {Jennewein}}, \bibinfo {author}
  {\bibfnamefont {T.}~\bibnamefont {Ursin}}, \ and\ \bibinfo {author}
  {\bibfnamefont {A.}~\bibnamefont {Zeilinger}},\ }\href@noop {} {\bibfield
  {journal} {\bibinfo  {journal} {Nature}\ }\textbf {\bibinfo {volume} {489}},\
  \bibinfo {pages} {269} (\bibinfo {year} {2012})}\BibitemShut {NoStop}%
\bibitem [{\citenamefont {Huang}\ \emph {et~al.}(2020)\citenamefont {Huang},
  \citenamefont {Huang},\ and\ \citenamefont {Li}}]{tel_network}%
  \BibitemOpen
  \bibfield  {author} {\bibinfo {author} {\bibfnamefont {N.-N.}\ \bibnamefont
  {Huang}}, \bibinfo {author} {\bibfnamefont {W.-H.}\ \bibnamefont {Huang}}, \
  and\ \bibinfo {author} {\bibfnamefont {C.-M.}\ \bibnamefont {Li}},\
  }\href@noop {} {\bibfield  {journal} {\bibinfo  {journal} {Scientific
  Reports}\ }\textbf {\bibinfo {volume} {10}} (\bibinfo {year}
  {2020})}\BibitemShut {NoStop}%
\bibitem [{\citenamefont {Deutsch}\ \emph {et~al.}(1996)\citenamefont
  {Deutsch}, \citenamefont {Ekert}, \citenamefont {Macchiavello}, \citenamefont
  {Popescu},\ and\ \citenamefont {Sanpera}}]{QP2Deutsch}%
  \BibitemOpen
  \bibfield  {author} {\bibinfo {author} {\bibfnamefont {D.}~\bibnamefont
  {Deutsch}}, \bibinfo {author} {\bibfnamefont {A.}~\bibnamefont {Ekert}},
  \bibinfo {author} {\bibfnamefont {C.}~\bibnamefont {Macchiavello}}, \bibinfo
  {author} {\bibfnamefont {S.}~\bibnamefont {Popescu}}, \ and\ \bibinfo
  {author} {\bibfnamefont {A.}~\bibnamefont {Sanpera}},\ }\href@noop {}
  {\bibfield  {journal} {\bibinfo  {journal} {Phys. Rev. Lett.}\ }\textbf
  {\bibinfo {volume} {77}},\ \bibinfo {pages} {2818} (\bibinfo {year}
  {1996})}\BibitemShut {NoStop}%
\bibitem [{\citenamefont {Dur}\ and\ \citenamefont {Briegel}(2003)}]{Ent_pur}%
  \BibitemOpen
  \bibfield  {author} {\bibinfo {author} {\bibfnamefont {W.}~\bibnamefont
  {Dur}}\ and\ \bibinfo {author} {\bibfnamefont {H.-J.}\ \bibnamefont
  {Briegel}},\ }\href@noop {} {\bibfield  {journal} {\bibinfo  {journal}
  {Phys.Rev. Lett.}\ }\textbf {\bibinfo {volume} {90}},\ \bibinfo {pages}
  {067901} (\bibinfo {year} {2003})}\BibitemShut {NoStop}%
\bibitem [{\citenamefont {Munro}\ \emph {et~al.}(2010)\citenamefont {Munro},
  \citenamefont {Harrison}, \citenamefont {Stephens}, \citenamefont {Devitt},\
  and\ \citenamefont {Nemoto}}]{Bill_2}%
  \BibitemOpen
  \bibfield  {author} {\bibinfo {author} {\bibfnamefont {W.~J.}\ \bibnamefont
  {Munro}}, \bibinfo {author} {\bibfnamefont {K.~A.}\ \bibnamefont {Harrison}},
  \bibinfo {author} {\bibfnamefont {A.~M.}\ \bibnamefont {Stephens}}, \bibinfo
  {author} {\bibfnamefont {S.~J.}\ \bibnamefont {Devitt}}, \ and\ \bibinfo
  {author} {\bibfnamefont {K.}~\bibnamefont {Nemoto}},\ }\href@noop {}
  {\bibfield  {journal} {\bibinfo  {journal} {Nature Photonics}\ }\textbf
  {\bibinfo {volume} {4}},\ \bibinfo {pages} {792} (\bibinfo {year}
  {2010})}\BibitemShut {NoStop}%
\bibitem [{\citenamefont {Ralph}\ \emph {et~al.}(2005)\citenamefont {Ralph},
  \citenamefont {Hayes},\ and\ \citenamefont {Gilchrist}}]{redundancy_code}%
  \BibitemOpen
  \bibfield  {author} {\bibinfo {author} {\bibfnamefont {T.~C.}\ \bibnamefont
  {Ralph}}, \bibinfo {author} {\bibfnamefont {A.~J.~F.}\ \bibnamefont {Hayes}},
  \ and\ \bibinfo {author} {\bibfnamefont {A.}~\bibnamefont {Gilchrist}},\
  }\href@noop {} {\bibfield  {journal} {\bibinfo  {journal} {Phys. Rev. Lett.}\
  }\textbf {\bibinfo {volume} {95}},\ \bibinfo {pages} {100501} (\bibinfo
  {year} {2005})}\BibitemShut {NoStop}%
\bibitem [{\citenamefont {Grassl}\ \emph {et~al.}(1999)\citenamefont {Grassl},
  \citenamefont {Geiselmann},\ and\ \citenamefont {Beth}}]{Reed_Sal}%
  \BibitemOpen
  \bibfield  {author} {\bibinfo {author} {\bibfnamefont {M.}~\bibnamefont
  {Grassl}}, \bibinfo {author} {\bibfnamefont {W.}~\bibnamefont {Geiselmann}},
  \ and\ \bibinfo {author} {\bibfnamefont {T.}~\bibnamefont {Beth}},\
  }\href@noop {} {\bibfield  {journal} {\bibinfo  {journal} {International
  Symposium on Applied Algebra, Algebraic Algorithms, and Error-Correcting
  Codes,}\ ,\ \bibinfo {pages} {231}} (\bibinfo {year} {1999})}\BibitemShut
  {NoStop}%
\bibitem [{\citenamefont {Fowler}\ \emph {et~al.}(2010)\citenamefont {Fowler},
  \citenamefont {Wang}, \citenamefont {Hill}, \citenamefont {Ladd},
  \citenamefont {Van~Meter},\ and\ \citenamefont {Hollenberg}}]{surfacecode}%
  \BibitemOpen
  \bibfield  {author} {\bibinfo {author} {\bibfnamefont {A.~G.}\ \bibnamefont
  {Fowler}}, \bibinfo {author} {\bibfnamefont {D.~S.}\ \bibnamefont {Wang}},
  \bibinfo {author} {\bibfnamefont {C.~D.}\ \bibnamefont {Hill}}, \bibinfo
  {author} {\bibfnamefont {T.~D.}\ \bibnamefont {Ladd}}, \bibinfo {author}
  {\bibfnamefont {R.}~\bibnamefont {Van~Meter}}, \ and\ \bibinfo {author}
  {\bibfnamefont {L.~C.~L.}\ \bibnamefont {Hollenberg}},\ }\href@noop {}
  {\bibfield  {journal} {\bibinfo  {journal} {Phys. Rev. Lett.}\ }\textbf
  {\bibinfo {volume} {104}},\ \bibinfo {pages} {180503} (\bibinfo {year}
  {2010})}\BibitemShut {NoStop}%
\bibitem [{\citenamefont {Gottesman}\ \emph {et~al.}(2001)\citenamefont
  {Gottesman}, \citenamefont {Kitaev},\ and\ \citenamefont {Preskill}}]{GKP1}%
  \BibitemOpen
  \bibfield  {author} {\bibinfo {author} {\bibfnamefont {D.}~\bibnamefont
  {Gottesman}}, \bibinfo {author} {\bibfnamefont {A.}~\bibnamefont {Kitaev}}, \
  and\ \bibinfo {author} {\bibfnamefont {J.}~\bibnamefont {Preskill}},\
  }\href@noop {} {\bibfield  {journal} {\bibinfo  {journal} {Phys. Rev. A}\
  }\textbf {\bibinfo {volume} {64}},\ \bibinfo {pages} {012310} (\bibinfo
  {year} {2001})}\BibitemShut {NoStop}%
\bibitem [{\citenamefont {Jiang}\ \emph {et~al.}(2009)\citenamefont {Jiang},
  \citenamefont {Taylor}, \citenamefont {Nemoto}, \citenamefont {Munro},
  \citenamefont {Van~Meter},\ and\ \citenamefont {Lukin}}]{Bill_3}%
  \BibitemOpen
  \bibfield  {author} {\bibinfo {author} {\bibfnamefont {L.}~\bibnamefont
  {Jiang}}, \bibinfo {author} {\bibfnamefont {J.~M.}\ \bibnamefont {Taylor}},
  \bibinfo {author} {\bibfnamefont {K.~M.}\ \bibnamefont {Nemoto}}, \bibinfo
  {author} {\bibfnamefont {W.~J.}\ \bibnamefont {Munro}}, \bibinfo {author}
  {\bibfnamefont {R.}~\bibnamefont {Van~Meter}}, \ and\ \bibinfo {author}
  {\bibfnamefont {M.~D.}\ \bibnamefont {Lukin}},\ }\href@noop {} {\bibfield
  {journal} {\bibinfo  {journal} {Phys. Rev. A}\ }\textbf {\bibinfo {volume}
  {79}},\ \bibinfo {pages} {032325} (\bibinfo {year} {2009})}\BibitemShut
  {NoStop}%
\bibitem [{\citenamefont {Cochrane}\ \emph {et~al.}(1999)\citenamefont
  {Cochrane}, \citenamefont {Milburn},\ and\ \citenamefont
  {Munro}}]{cat_states}%
  \BibitemOpen
  \bibfield  {author} {\bibinfo {author} {\bibfnamefont {P.~T.}\ \bibnamefont
  {Cochrane}}, \bibinfo {author} {\bibfnamefont {G.~J.}\ \bibnamefont
  {Milburn}}, \ and\ \bibinfo {author} {\bibfnamefont {W.~J.}\ \bibnamefont
  {Munro}},\ }\href@noop {} {\bibfield  {journal} {\bibinfo  {journal} {Phys.
  Rev. A}\ }\textbf {\bibinfo {volume} {59}},\ \bibinfo {pages} {2631}
  (\bibinfo {year} {1999})}\BibitemShut {NoStop}%
\bibitem [{\citenamefont {Michael}\ \emph {et~al.}(2016)\citenamefont
  {Michael}, \citenamefont {Silveri}, \citenamefont {Brierley}, \citenamefont
  {Albert}, \citenamefont {Salmilehto}, \citenamefont {Jiang},\ and\
  \citenamefont {Girvin}}]{binomial_code}%
  \BibitemOpen
  \bibfield  {author} {\bibinfo {author} {\bibfnamefont {M.~H.}\ \bibnamefont
  {Michael}}, \bibinfo {author} {\bibfnamefont {M.}~\bibnamefont {Silveri}},
  \bibinfo {author} {\bibfnamefont {R.~T.}\ \bibnamefont {Brierley}}, \bibinfo
  {author} {\bibfnamefont {V.~V.}\ \bibnamefont {Albert}}, \bibinfo {author}
  {\bibfnamefont {J.}~\bibnamefont {Salmilehto}}, \bibinfo {author}
  {\bibfnamefont {L.}~\bibnamefont {Jiang}}, \ and\ \bibinfo {author}
  {\bibfnamefont {S.~M.}\ \bibnamefont {Girvin}},\ }\href@noop {} {\bibfield
  {journal} {\bibinfo  {journal} {Phys. Rev. X}\ }\textbf {\bibinfo {volume}
  {6}},\ \bibinfo {pages} {031006} (\bibinfo {year} {2016})}\BibitemShut
  {NoStop}%
\bibitem [{\citenamefont {Azuma}\ and\ \citenamefont
  {Kato}(2017)}]{quantum_capt}%
  \BibitemOpen
  \bibfield  {author} {\bibinfo {author} {\bibfnamefont {K.}~\bibnamefont
  {Azuma}}\ and\ \bibinfo {author} {\bibfnamefont {G.}~\bibnamefont {Kato}},\
  }\href@noop {} {\bibfield  {journal} {\bibinfo  {journal} {Phys. Rev. A}\
  }\textbf {\bibinfo {volume} {96}},\ \bibinfo {pages} {032332} (\bibinfo
  {year} {2017})}\BibitemShut {NoStop}%
\bibitem [{\citenamefont {Shirokov}(2017)}]{q_capacity3}%
  \BibitemOpen
  \bibfield  {author} {\bibinfo {author} {\bibfnamefont {M.~E.}\ \bibnamefont
  {Shirokov}},\ }\href@noop {} {\bibfield  {journal} {\bibinfo  {journal}
  {Journal of Mathematical Physics}\ }\textbf {\bibinfo {volume} {58}},\
  \bibinfo {pages} {102202} (\bibinfo {year} {2017})}\BibitemShut {NoStop}%
\bibitem [{\citenamefont {Rosati}\ \emph {et~al.}(2018)\citenamefont {Rosati},
  \citenamefont {Mari},\ and\ \citenamefont {Giovannetti}}]{q_capacity2}%
  \BibitemOpen
  \bibfield  {author} {\bibinfo {author} {\bibfnamefont {M.}~\bibnamefont
  {Rosati}}, \bibinfo {author} {\bibfnamefont {A.}~\bibnamefont {Mari}}, \ and\
  \bibinfo {author} {\bibfnamefont {V.}~\bibnamefont {Giovannetti}},\
  }\href@noop {} {\bibfield  {journal} {\bibinfo  {journal} {Nat. Commun.}\
  }\textbf {\bibinfo {volume} {9}},\ \bibinfo {pages} {4339} (\bibinfo {year}
  {2018})}\BibitemShut {NoStop}%
\bibitem [{\citenamefont {Pirandola}(2019{\natexlab{a}})}]{stefano1}%
  \BibitemOpen
  \bibfield  {author} {\bibinfo {author} {\bibfnamefont {S.}~\bibnamefont
  {Pirandola}},\ }\href@noop {} {\bibfield  {journal} {\bibinfo  {journal}
  {Commun. Phys}\ }\textbf {\bibinfo {volume} {2}},\ \bibinfo {pages} {51}
  (\bibinfo {year} {2019}{\natexlab{a}})}\BibitemShut {NoStop}%
\bibitem [{\citenamefont {Pirandola}(2019{\natexlab{b}})}]{stefano2}%
  \BibitemOpen
  \bibfield  {author} {\bibinfo {author} {\bibfnamefont {S.}~\bibnamefont
  {Pirandola}},\ }\href@noop {} {\bibfield  {journal} {\bibinfo  {journal}
  {Quantum Science and Technology}\ }\textbf {\bibinfo {volume} {4}} (\bibinfo
  {year} {2019}{\natexlab{b}})}\BibitemShut {NoStop}%
\bibitem [{\citenamefont {Fanizza}\ \emph {et~al.}(2020)\citenamefont
  {Fanizza}, \citenamefont {Kianvash},\ and\ \citenamefont
  {Giovannetti}}]{q_capacity1}%
  \BibitemOpen
  \bibfield  {author} {\bibinfo {author} {\bibfnamefont {M.}~\bibnamefont
  {Fanizza}}, \bibinfo {author} {\bibfnamefont {F.}~\bibnamefont {Kianvash}}, \
  and\ \bibinfo {author} {\bibfnamefont {V.}~\bibnamefont {Giovannetti}},\
  }\href@noop {} {\bibfield  {journal} {\bibinfo  {journal} {Phys. Rev. Lett.}\
  }\textbf {\bibinfo {volume} {125}},\ \bibinfo {pages} {020503} (\bibinfo
  {year} {2020})}\BibitemShut {NoStop}%
\bibitem [{\citenamefont {Ralph}\ \emph {et~al.}(2007)\citenamefont {Ralph},
  \citenamefont {Resch},\ and\ \citenamefont {Gilchrist}}]{toffoli1}%
  \BibitemOpen
  \bibfield  {author} {\bibinfo {author} {\bibfnamefont {T.~C.}\ \bibnamefont
  {Ralph}}, \bibinfo {author} {\bibfnamefont {K.~J.}\ \bibnamefont {Resch}}, \
  and\ \bibinfo {author} {\bibfnamefont {A.}~\bibnamefont {Gilchrist}},\
  }\href@noop {} {\bibfield  {journal} {\bibinfo  {journal} {Phys. Rev. A}\
  }\textbf {\bibinfo {volume} {75}},\ \bibinfo {pages} {022313} (\bibinfo
  {year} {2007})}\BibitemShut {NoStop}%
\bibitem [{\citenamefont {Ioniciouiu}\ \emph {et~al.}(2009)\citenamefont
  {Ioniciouiu}, \citenamefont {Spiller},\ and\ \citenamefont
  {Munro}}]{toffoli2}%
  \BibitemOpen
  \bibfield  {author} {\bibinfo {author} {\bibfnamefont {R.}~\bibnamefont
  {Ioniciouiu}}, \bibinfo {author} {\bibfnamefont {T.~P.}\ \bibnamefont
  {Spiller}}, \ and\ \bibinfo {author} {\bibfnamefont {W.~J.}\ \bibnamefont
  {Munro}},\ }\href@noop {} {\bibfield  {journal} {\bibinfo  {journal} {Phys.
  Rev. A}\ }\textbf {\bibinfo {volume} {80}},\ \bibinfo {pages} {012312}
  (\bibinfo {year} {2009})}\BibitemShut {NoStop}%
\bibitem [{\citenamefont {Lo~Piparo}\ \emph {et~al.}(2020)\citenamefont
  {Lo~Piparo}, \citenamefont {Hanks}, \citenamefont {Gravel}, \citenamefont
  {Munro},\ and\ \citenamefont {Nemoto}}]{QuM}%
  \BibitemOpen
  \bibfield  {author} {\bibinfo {author} {\bibfnamefont {N.}~\bibnamefont
  {Lo~Piparo}}, \bibinfo {author} {\bibfnamefont {M.}~\bibnamefont {Hanks}},
  \bibinfo {author} {\bibfnamefont {C.}~\bibnamefont {Gravel}}, \bibinfo
  {author} {\bibfnamefont {W.~J.}\ \bibnamefont {Munro}}, \ and\ \bibinfo
  {author} {\bibfnamefont {K.}~\bibnamefont {Nemoto}},\ }\href@noop {}
  {\bibfield  {journal} {\bibinfo  {journal} {Phys. Rev. Lett.}\ }\textbf
  {\bibinfo {volume} {124}},\ \bibinfo {pages} {210503} (\bibinfo {year}
  {2020})}\BibitemShut {NoStop}%
\bibitem [{\citenamefont {Kwiat}(1997)}]{hyperentanglement1}%
  \BibitemOpen
  \bibfield  {author} {\bibinfo {author} {\bibfnamefont {P.~G.}\ \bibnamefont
  {Kwiat}},\ }\href@noop {} {\bibfield  {journal} {\bibinfo  {journal} {J. Mod.
  Opt.}\ }\textbf {\bibinfo {volume} {44}},\ \bibinfo {pages} {2173} (\bibinfo
  {year} {1997})}\BibitemShut {NoStop}%
\bibitem [{\citenamefont {Barreiro}\ \emph {et~al.}(2005)\citenamefont
  {Barreiro}, \citenamefont {Langford}, \citenamefont {Peters},\ and\
  \citenamefont {Kwiat}}]{hyper2}%
  \BibitemOpen
  \bibfield  {author} {\bibinfo {author} {\bibfnamefont {J.~T.}\ \bibnamefont
  {Barreiro}}, \bibinfo {author} {\bibfnamefont {N.~K.}\ \bibnamefont
  {Langford}}, \bibinfo {author} {\bibfnamefont {N.~A.}\ \bibnamefont
  {Peters}}, \ and\ \bibinfo {author} {\bibfnamefont {P.~G.}\ \bibnamefont
  {Kwiat}},\ }\href@noop {} {\bibfield  {journal} {\bibinfo  {journal} {Phys.
  Rev. Lett.}\ }\textbf {\bibinfo {volume} {95}},\ \bibinfo {pages} {1}
  (\bibinfo {year} {2005})}\BibitemShut {NoStop}%
\bibitem [{\citenamefont {Wei}\ \emph {et~al.}(2007)\citenamefont {Wei},
  \citenamefont {Barreiro},\ and\ \citenamefont {Kwiat}}]{hyper3}%
  \BibitemOpen
  \bibfield  {author} {\bibinfo {author} {\bibfnamefont {T.~C.}\ \bibnamefont
  {Wei}}, \bibinfo {author} {\bibfnamefont {J.~T.}\ \bibnamefont {Barreiro}}, \
  and\ \bibinfo {author} {\bibfnamefont {P.~G.}\ \bibnamefont {Kwiat}},\
  }\href@noop {} {\bibfield  {journal} {\bibinfo  {journal} {Phys. Rev. A}\
  }\textbf {\bibinfo {volume} {75}},\ \bibinfo {pages} {1} (\bibinfo {year}
  {2007})}\BibitemShut {NoStop}%
\bibitem [{\citenamefont {Muralidharan}\ \emph {et~al.}(2018)\citenamefont
  {Muralidharan}, \citenamefont {Zou}, \citenamefont {Li},\ and\ \citenamefont
  {Jiang}}]{PS_QRS}%
  \BibitemOpen
  \bibfield  {author} {\bibinfo {author} {\bibfnamefont {S.}~\bibnamefont
  {Muralidharan}}, \bibinfo {author} {\bibfnamefont {C.-L.}\ \bibnamefont
  {Zou}}, \bibinfo {author} {\bibfnamefont {L.}~\bibnamefont {Li}}, \ and\
  \bibinfo {author} {\bibfnamefont {L.}~\bibnamefont {Jiang}},\ }\href@noop {}
  {\bibfield  {journal} {\bibinfo  {journal} {Phys. Rev. A}\ }\textbf {\bibinfo
  {volume} {97}},\ \bibinfo {pages} {052316} (\bibinfo {year}
  {2018})}\BibitemShut {NoStop}%
\bibitem [{\citenamefont {Aliferis}\ \emph {et~al.}(2006)\citenamefont
  {Aliferis}, \citenamefont {Gottesman},\ and\ \citenamefont
  {Preskill}}]{ftolerant_threshold}%
  \BibitemOpen
  \bibfield  {author} {\bibinfo {author} {\bibfnamefont {P.}~\bibnamefont
  {Aliferis}}, \bibinfo {author} {\bibfnamefont {D.}~\bibnamefont {Gottesman}},
  \ and\ \bibinfo {author} {\bibfnamefont {J.}~\bibnamefont {Preskill}},\
  }\href@noop {} {\bibfield  {journal} {\bibinfo  {journal} {Quant. Inf.
  Comput.}\ }\textbf {\bibinfo {volume} {6}},\ \bibinfo {pages} {97} (\bibinfo
  {year} {2006})}\BibitemShut {NoStop}%
\end{thebibliography}%

\appendix

\section{The $[[3,1,2]]_{3}$ QRS code}

Here we derive Eq. \eqref{eq:logic_qutrit} by using the logic scheme
of Fig. \ref{fig:logic_qutrit_qubit}(a). We begin as shown in the
Figure by initializing the three qutrits as $\left.|\psi\right\rangle _{1}=\left.\alpha|0\right\rangle +\left.\beta|1\right\rangle +\left.\gamma|2\right\rangle ,$
$\left.|\psi\right\rangle _{2}=\left.|0\right\rangle $ and $\left.|\psi\right\rangle _{3}=\left.|0\right\rangle +\left.|1\right\rangle +\left.|2\right\rangle ,$
respectively. We then apply the CX gate represented in Fig. \ref{fig:logic_qutrit_qubit}(a)
with a red box. This gate will perform the following operations on
the qutrits: $\left.CX|0\right\rangle \left.|0\right\rangle =\left.|0\right\rangle \left.|0\right\rangle ,$
$\left.CX|1\right\rangle \left.|0\right\rangle =\left.|1\right\rangle \left.|2\right\rangle $
and $\left.CX|2\right\rangle \left.|0\right\rangle =\left.|2\right\rangle \left.|1\right\rangle ,$
respectively. 

We then apply the control cyclic gate between qutrit 3 and 2 and between
3 and 1, respectively. This gate adds mod-2 $n$ units to the target
qutrit, where $n$ is the state of the control qutrit. The resulting
state is expressed by Eq. \ref{eq:logic_qutrit}.

\section{The qubits encoding of the $[[3,1,2]]_{3}$ QRS code}

\begin{figure}
\begin{centering}
\includegraphics[scale=0.4]{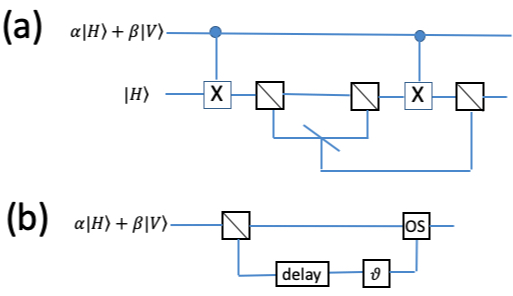}
\par\end{centering}
\caption{\label{fig:initial_states}Logic circuit to create a qutrit using
(a) two polarized photons and (b) one multiplexed photon.}
\end{figure}

In this Appendix\textcolor{black}{{} we show how we can construct the
$[[3,1,2]]_{3}$ QRS code using six polarized photons. It is illustrative
to begin by showing how one can encode a single qutrit into to two
polarization encoded photons (see Fig. \ref{fig:initial_states}(a)).}
Our initial state is of the form\textcolor{black}{{} }$\left.|\psi\right\rangle _{1}=\alpha\left.|0\right\rangle _{1}+\beta\left.|1\right\rangle _{1}$
and $\left.|\psi\right\rangle _{2}=\left.|0\right\rangle _{2}$, where
$\left.|0\right\rangle =\left.|H\right\rangle $ and $\left.|1\right\rangle =\left.|V\right\rangle .$
After applying a CNOT gate and letting the second photon pass through
a polarizing beam splitter (PBS) we have the two-qubit entangled state
$\alpha\left.|00\right\rangle +\beta\left.|11\right\rangle .$ We
now apply a second PBS to photon 2 followed by the letting the V-component
pass through an unbalanced BS which add a phase $\vartheta$ to this
component. The resulting state is of the form $\alpha\left.|00\right\rangle +\beta\left.\cos\vartheta|11\right\rangle \left.|0\right\rangle +\beta\left.\sin\vartheta|10\right\rangle \left.|1\right\rangle .$
The components are recombined into the same initial modes by applying
two PBSs and a CNOT to give $\alpha\left.|00\right\rangle +\beta\left.\cos\vartheta|10\right\rangle +\beta\left.\sin\vartheta|11\right\rangle .$
Finally, through a swap gate we obtain the desired initial state 

\noindent 
\begin{equation}
\alpha\left.|00\right\rangle +\beta\left.\cos\vartheta|01\right\rangle +\beta\left.\sin\vartheta|11\right\rangle .\label{eq:initial_qutrit_qubits}
\end{equation}

With this initial state we can now apply the logic circuit of Fig.
\ref{fig:logic_qutrit_qubit}(b) to create deterministically the desired
logic qutrit $\left.|D\right\rangle $ 

\noindent 
\begin{equation}
\begin{array}{c}
\begin{array}{c}
\begin{array}{c}
\left.|0\right\rangle _{L}=\left.|H_{1}H_{2}H_{3}H_{4}H_{5}H_{6}\right\rangle +\left.|V_{1}H_{2}V_{3}H_{4}V_{5}H_{6}\right\rangle \end{array}\\
\left.+|V_{1}V_{2}V_{3}V_{4}V_{5}V_{6}\right\rangle ,\\
\left.|1\right\rangle _{L}=\left.|H_{1}H_{2}V_{3}H_{4}V_{5}V_{6}\right\rangle +\left.|V_{1}H_{2}V_{3}V_{4}H_{5}H_{6}\right\rangle \\
+\left.|V_{1}V_{2}H_{3}H_{4}V_{5}H_{6}\right\rangle ,\\
\left.|2\right\rangle _{L}=\left.|H_{1}H_{2}V_{3}V_{4}V_{5}H_{6}\right\rangle +\left.|V_{1}H_{2}H_{3}H_{4}V_{5}V_{6}\right\rangle \\
+\left.|V_{1}V_{2}V_{3}H_{4}H_{5}H_{6}\right\rangle .
\end{array}\end{array}\label{eq:logic_qutrits_6ph}
\end{equation}

\section{Quantum multiplexing applied to the $[[3,1,2]]_{3}$ QRS code}

One can consider using 3 quantum multiplexed photons (instead of 6
photons) each carrying 2 qubits to encode the $[[3,1,2]]_{3}$ QRS
QEC. In this case we only need 3 photons as the total number of qubits
will be the same as in the non-multiplexing case. For the second qubit
we use the time-bin degree of freedom, having $S$ and $L$ as the
short and long component, respectively. Accordingly, the component
of the second photon required to encode one qutrit can be expressed
by the time-bin components as $H\rightarrow S$ and $V\rightarrow L.$
The initial state can be created by using the logic circuit of Fig.
\ref{fig:initial_states}(b). In this case we start with a single
photon given by $\left.|\psi\right\rangle =\alpha\left.|H\right\rangle +\beta\left.|V\right\rangle .$
Applying a PBS and a delay followed by a $\vartheta$ rotation on
the V component will give $\alpha\left.|H_{S}\right\rangle +\beta\left.\cos\vartheta|H_{L}\right\rangle +\beta\left.\sin\vartheta|V_{L}\right\rangle .$
An optical switch (OS) will recombine the components into the same
mode to create the initial state. 
\begin{table}
\begin{centering}
\begin{tabular}{|c|c|c|}
\hline 
Control & Target & Procedure\tabularnewline
\hline 
\hline 
Polarization & Polarization & Atomic interaction\tabularnewline
\hline 
Polarization & Time-bin & %
\begin{tabular}{|c|}
\hline 
Swap $H_{S_{2}}$ with $V_{L_{2}}$ \tabularnewline
\hline 
\hline 
CNOT Pol.1 Pol.2\tabularnewline
\hline 
Swap $H_{S_{2}}$ with $V_{L_{2}}$ \tabularnewline
\hline 
\end{tabular}\tabularnewline
\hline 
Time-bin  & Polarization & %
\begin{tabular}{|c|}
\hline 
Swap $H_{L_{1}}$ with $V_{S_{1}}$ \tabularnewline
\hline 
\hline 
CNOT Pol.1 Pol.2\tabularnewline
\hline 
Swap $H_{L_{1}}$ with $V_{S_{1}}$ \tabularnewline
\hline 
\end{tabular}\tabularnewline
\hline 
Time-bin & Time-bin & %
\begin{tabular}{|c|}
\hline 
Swap $H_{L_{1}}$ with $V_{S_{1}}$ \tabularnewline
\hline 
\hline 
CNOT Pol.1 TB2\tabularnewline
\hline 
Swap $H_{L_{1}}$ with $V_{S_{1}}$ \tabularnewline
\hline 
\end{tabular}\tabularnewline
\hline 
\end{tabular}
\par\end{centering}
\caption{\label{tab:tableDOF}The operations required to perform a CNOT gate
between different DOFs of two quantum multiplexed photons. }
\end{table}

In this case the corresponding logic qutrit based states of Eq. \ref{eq:logic_qutrits_6ph}
will be

\noindent 
\[
\begin{array}{c}
\begin{array}{c}
\left.|0\right\rangle _{L}=\left.|H_{S_{1}}H_{S_{2}}H_{S_{3}}\right\rangle +\left.|V_{S_{1}}V_{S_{2}}V_{S_{3}}\right\rangle \left.+|V_{L_{1}}V_{L_{2}}V_{L_{3}}\right\rangle \\
\left.|1\right\rangle _{L}=\left.|H_{S_{1}}V_{S_{2}}V_{L_{3}}\right\rangle +\left.|V_{S_{`1}}V_{L_{2}}H_{S_{3}}\right\rangle +\left.|V_{L_{!}}H_{S_{2}}V_{S_{3}}\right\rangle \\
\left.|2\right\rangle _{L}=\left.|H_{S_{1}}V_{L_{2}}V_{S_{3}}\right\rangle +\left.|V_{S_{1}}H_{S_{2}}V_{L_{3}}\right\rangle +\left.|V_{L_{1}}V_{S_{2}}H_{S_{3}}\right\rangle 
\end{array}\end{array}
\]

The basic gates that allow to reproduce entirely the corresponding
logic gates of Fig. \ref{fig:logic_qutrit_qubit}(b) are CNOT gates
between the polarization degrees of freedom (DOFs) (which we assume
it can be mediated by an atom), between polarization DOF and time-bin
DOF and between time-bin and time-bin DOF of two quantum multiplexed
photons. Table \ref{tab:tableDOF} summarizes the procedure used to
perform a deterministic CNOT gate between two generic DOFs of photon
$1$ (Control) and photon $2$ (Target), respectively. The swapping
of the components can be performed by using linear optical elements
are shown in the Supplemental material of \cite{QuM}. Finally, the
Toffoli gates in Fig. \ref{fig:logic_qutrit_qubit}(b) can be decomposed
as a series of single qubits rotation and a single CNOT gate between
the DOFs given in Table \ref{tab:tableDOF}.

\section{High dimensional QRS code}

In this Appendix we discuss the advantages arising from applying the
aggregate network technique to the quantum multiplexing 211-qudit
QRS code where each photon is carrying 8 qubits. The results for such
a code are illustrated in Figure \ref{fig:first_level_graph}(a).
Here the black dotted curve is the transmission probability required
to reach $\overline{P}_{S}$ when all qudits are traveling in the
same channel. The colored curves correspond to the aggregate network
case where 50 (orange), 80 (yellow) and 100 (purple) qudits are traveling
in the higher transmitivity channel. 
\end{document}